\documentclass[final,1p,times]{elsarticle}

\usepackage[utf8]{inputenc}
\usepackage[english]{babel}
\usepackage{amsmath, amssymb, amsfonts}
\usepackage{eqnarray}
\usepackage{graphicx}
\usepackage{subfigure}
\usepackage{tabularx}
\usepackage{cleveref}
\usepackage{soul}
\usepackage{xcolor}
\usepackage{lipsum}
\usepackage{xargs}

\newcommand{\mihi}[1]{\textcolor{black}{#1}}
\newcommand{\rev}[1]{\textcolor{black}{#1}}
\newcommand{\Lorenzo}[1]{{\color{black} #1}}
\newcounter{lnote}

\newcommand{\bbeta}{\boldsymbol{\beta}}

\newcommand{\jj}{\mathbf{j}}
\newcommand{\vv}{\mathbf{v}}
\newcommand{\xx}{\mathbf{x}}

\newcommand{\qoiscal}{f}
\newcommand{\mcI}{\mathcal{I}}
\newcommand{\mcS}{\mathcal{S}}
\newcommand{\mcU}{\mathcal{U}}
\newcommand{\s}{\, \, \, \,\,}

\DeclareMathOperator*{\argmax}{arg\,max}
\DeclareMathOperator*{\argmin}{arg\,min}

\begin{document}

	\title{\Lorenzo{Data-informed uncertainty quantification for laser-based powder bed fusion additive manufacturing}}
	
\author[UniPV]{Mihaela Chiappetta}
\author[TUM]{Chiara Piazzola} 
\author[IMATI]{Lorenzo Tamellini}
\author[UniPV,IMATI]{Alessandro Reali}
\author[UniPV,IMATI,IRCCS]{Ferdinando Auricchio}
\author[UniPV]{Massimo Carraturo}
\address[UniPV]{Department of Civil Engineering and Architecture - University of Pavia, Via Ferrata 3, 27100, Pavia, Italy}
\address[TUM]{Department of Mathematics - Technical University of Munich, Boltzmannstra\ss e, 3, 85748, Garching bei M\"unchen, Germany}  
\address[IMATI]{Istituto di Matematica Applicata e Tecnologie Informatiche “E. Magenes” - Consiglio Nazionale delle Ricerche, Via Ferrata, 5/A, 27100, Pavia, Italy}  
\address[IRCCS]{Fondazione IRCCS Policlinico San Matteo, 27100 Pavia, Italy}
	

	\begin{abstract}
		We present an efficient approach to quantify the uncertainties associated with the numerical simulations of the laser-based powder bed fusion of metals processes.    
	Our study focuses on a thermomechanical model of an Inconel 625 cantilever beam, based on the AMBench2018-01 benchmark proposed by the National Institute of Standards and Technology (NIST). 
	The proposed approach consists of a \Lorenzo{forward uncertainty quantification analysis of the residual strains of the cantilever beam given the uncertainty in some of the parameters of the numerical simulation, namely the powder convection coefficient and the activation temperature. The uncertainty on such parameters is modelled by a
		data-informed probability density function obtained by a Bayesian inversion procedure, based on the displacement experimental data provided by NIST.
		To overcome the computational challenges of both the Bayesian inversion and the forward uncertainty quantification analysis we employ
		a multi-fidelity surrogate modelling technique, specifically the multi-index stochastic collocation method.
		The proposed approach allows us to achieve a 33\% reduction in the uncertainties on the prediction of residual strains
		compared with what we would get basing the forward UQ analysis on a-priori ranges for the uncertain parameters,
		and in particular the mode of the probability density function of such quantities (i.e., its ``most likely value'', roughly speaking)
		results to be in good agreement with the experimental data provided by NIST,
		even though only displacement data were used for the Bayesian inversion procedure.}
	\end{abstract}

	
		\maketitle
	
	\renewcommand\thefootnote{\fnsymbol{footnote}}
	\setcounter{footnote}{1}

	\section{Introduction}
	\label{sec:intro}
	Laser-Based Powder Bed Fusion of Metals (PBF-LB/M) \cite{bhavar2017review,king2015laser} is an additive manufacturing process
	\cite{gibson2021additive,wong2012review} in which a laser beam selectively melts a layer of metal powder and builds the final metal part using a layer-by-layer strategy. 
	PBF-LB/M is a versatile and precise technology capable of producing complex geometries and structures with high resolution and accuracy.  
	It is also capable of producing high-strength metal parts with excellent mechanical properties, making it a popular choice for applications in the aerospace, automotive, medical and other industries. 
	In addition, the process allows the use of a wide range of metal powders, including titanium, stainless steel, aluminum and other alloys, making it possible to produce metal parts with a large variety of properties. 
	However, the limited knowledge of the complex physical phenomena involved in PBF-LB/M and the sources of uncertainty due to powder particle properties, process parameters, machine settings, thermal effects and environmental conditions, can compromise the quality of the manufactured metal parts \cite{king2015laser,lopez2016identifying}. 
	A key to addressing the above uncertainties and limited process knowledge is the application of Uncertainty Quantification (UQ) methodologies \cite{smith2013uncertainty,ghanem2017handbook,soize2017uncertainty}. 
	
	\Lorenzo{Such methodologies aim at assessing how the uncertainties, e.g., on material properties and process parameters, affect the results of a numerical simulation of the PBF-LB/M process, or, in other words, how the uncertainties on the model inputs propagates to the solution/model outputs.
		For example, a classical goal of UQ is to compute the first statistical moments (expected values, standard deviation, skewness, kurtosis)
		of the model outputs, or ideally their probability density function (PDF), from which information on the statistical moments can be easily determined.}
	The UQ for PBF-LB/M thus provides a robust framework for understanding and communicating the impact of the various sources of uncertainty
	that occur during the process, enabling informed decision making and improved optimisation of the PBF-LB/M process.
	
	The research community has \Lorenzo{investigated} 
	several UQ techniques for PBF-LB/M processes 
	\cite{ahu2017uncertainty,hu2017uncertainty,mahadevan2022uncertainty}. 
	However, the application of UQ to PBF-LB/M numerical models \Lorenzo{still} presents challenges,
	mainly related to the required number of numerical simulations and the associated computational costs. 
	\Lorenzo{Efficient algorithms and high performance computing resources are then needed
		to manage the computational complexity and to make the UQ process feasible for practical implementation.
		In this context, surrogate modelling is usually integrated into the UQ for PBF-LB/M numerical models.}
	
	Surrogate modelling involves the construction of response surfaces, also known as surrogate models,
	through interpolation or approximation of the results obtained from a limited number of numerical simulations
	\Lorenzo{of the PBF-LB/M process for judiciously selected values of the uncertain parameters, see \cite{sudret2017surrogate} for a general overview}. 
	\rev{Once a surrogate model is built, its evaluations (which are typically very cheap to obtain) can replace the evaluations of the full physics-based model in the UQ analysis, with great computational savings. Of course, the number of simulations needed to build reliable surrogate models is problem-dependent, and in certain situations such number might be so large that a surrogate-model-based approach is no longer interesting (roughly speaking, this might happen, e.g., when the application depends on a very large number of parameters, or when the model outputs depends non-smoothly on the model inputs). However, for a wide range of situations, accurate surrogate models can be built at a reasonable cost, and thus represent an efficient and cheaper alternative to a purely sampling-based approach, allowing rapid evaluation of statistical properties (moments, PDF) of the model outputs.}
	The following are examples of surrogate modelling for the application of the UQ to PBF-LB/M numerical models.
	Wang et al. \cite{wang2020uncertainty} propose a sequential Bayesian calibration method to perform experimental parameter calibration and model correction
	to significantly improve the validity of the melt pool surrogate model. 
	Nath et al. \cite{nath2017mutli} present an inverse UQ framework to predict microstructure evolution during solidification
	by coupling a meso-scale melt pool surrogate model with a micro-scale cellular automata model. 
	Ghosh et al. \cite{ghosh2019uncertainty} use surrogate models to quantify the contribution of different sources of uncertainty
	to micro-segregation variability during PBF-LB/M processes. 
	Kamath \cite{kamath2016data} uses feature selection techniques to identify important variables, data-driven surrogate models to reduce computational cost,
	improved sampling techniques to adequately cover the design space, and uncertainty analysis for statistical inference. \rev{Surrogate-based modelling is also a valuable tool for optimising PBF-LB/M processes, as it provides an effective vehicle for improving overall process efficiency, again by providing cheap approximations that can efficiently replace the expensive evaluations of the full physics-based model. Several surrogate-based optimisation approaches have been successfully implemented in numerical simulations of various additive manufacturing processes including PBF-LB/M \cite{goguelin2021bayesian,wang2023optimization,zhang2021bayesian}. Tamellini et al. \cite{tamellini2020parametric} emphasised the effectiveness of using surrogate models for parametric optimisation in the field of combined additive-subtractive manufacturing.}
	
	\Lorenzo{
		In the present manuscript, we adopt an efficient approach to quantify the uncertainties associated with numerical simulations of the PBF-LB/M process. 
		We consider a thermomechanical Inconel 625 cantilever beam model based on the AMBench2018-01
		benchmark proposed by the National Institute of Standards and Technology (NIST) \cite{NIST}.
		In particular, our goal is to perform a forward UQ analysis of the residual strains of the cantilever beam
		given the uncertainty on some of the parameters of the numerical simulation, namely the
		powder convection coefficient and the activation temperature;
		these two parameters were found to be the most influential on the outputs of the numerical simulation in
		our previous work \cite{chiappetta2022inverse}. The uncertainty on such parameters is modelled by a
		data-informed PDF obtained by a Bayesian inversion procedure
		\cite{box2011bayesian,berger2013statistical,zhang2003compound,stuart:acta.bayesian,thanh-bui.gattas:MCMC}
		based on the displacement experimental data provided by NIST.
		To overcome the computational challenges of both the Bayesian inversion and the following forward UQ analysis we employ
		the Multi-Index Stochastic Collocation (MISC) multi-fidelity surrogate modelling technique
		\cite{beck2019iga,hajiali.eal:MISC1,jakeman2019adaptive,piazzola.eal:ferry-paper}, 
		a methodology that has not been previously applied to UQ for PBF-LB/M processes  to the best of our knowledge.}
	
	\Lorenzo{Multi fidelity is a surrogate modelling paradigm by which most of the variability of the model outputs due
		to the uncertainty on the parameters is explored by resorting at first to low-resolution numerical approximations of the PBF-LB/M process only;
		once the main trends of the model outputs with respect to the parameters are captured,
		a few runs of the high-resolution numerical approximations of the PBF-LB/M process are then incorporated in the surrogate model,
		correcting any bias introduced in the previous step.
		Since this procedure resorts only sparingly to high-resolution approximations,
		the overall cost is substantially smaller than what it would take to build a surrogate model querying the high-resolution approximations only,
		yet maintaining a good accuracy.} 
	\Lorenzo{As already hinted, the current study is an extension of our previous work \cite{chiappetta2022inverse},
		in which a similar workflow was carried out, relying however on a single-fidelity surrogate modelling technique, namely sparse grids
		\cite{babuska.nobile.eal:stochastic2,xiu.hesthaven:high,piazzola2022sparse,b.griebel:acta}.}
	
	\Lorenzo{
		The results of the forward UQ analysis are encouraging: in particular, basing the forward UQ on the
		data-informed PDF of the uncertain parameters computed by the Bayesian inversion allows us to achieve
		a 33\% reduction in the uncertainties on the prediction of residual strains compared to what we
		would get basing the forward UQ analysis on a-priori ranges for the uncertain parameters.
		In particular the mode of the PDF of the residual strains (i.e., their \lq\lq most likely
		value", roughly speaking) results to be in good agreement with the experimental data provided
		by NIST, even though only displacement data were used in the Bayesian inversion procedure.}
	
	The structure of the present work is as follows. 
	In \Cref{sec:governinEquations}, we present the governing equations describing the thermal and mechanical problems involved in the PBF-LB/M process. 
	In \Cref{sec:ExperimentalSetup}, we describe the AMBench2018-01 benchmark performed by NIST. 
	In \Cref{sec:numericalSimulations}, we describe the PBF-LB/M numerical simulations. 
	In \Cref{sec:MFSG}, we present our UQ approach based on multi-fidelity surrogate models. 
	In \Cref{sec:resultsDiscussion}, we report and discuss the numerical results. 
	Finally, in \Cref{sec:Conclusion}, we draw the main conclusions and possible further perspectives of the present work.
	
	\section{Governing equations}
	\label{sec:governinEquations}
	
	In the present study, we use a part-scale thermomechanical finite element model to simulate the PBF-LB/M process.
	In this section, we describe the governing equations of the thermal and mechanical problem for our study.
	For further details on thermomechanical finite element modelling, we refer the reader to \cite{chiumenti2017numerical,denlinger2017thermomechanical,carraturo2020modeling,carraturo2020numerical,li2018modeling,tan2019thermo,carraturo2022two,carraturo2021immersed,king2015overview, schoinochoritis2017simulation}.

	\subsection{Thermal problem}
	\label{subsec:Th}
	To solve the thermal problem of the PBF-LB/M process, we adopt a transient thermal analysis.
	In particular, assuming that the material follows Fourier's law, the thermal problem is governed by the following temperature dependent heat transfer equation:
	\begin{equation*}
		\rho c_p(T) \dfrac{\partial T}{\partial t} -  \nabla \cdot \left(k(T) \nabla T\right)=0 \s \mathrm{in} \s \Omega,
		\centering
		\label{eq: eq1}
	\end{equation*}
	where $T$ is the temperature field, $\rho$ denotes the constant material density, $c_p$ is the temperature dependent specific heat capacity at constant pressure, and $k$ is the temperature dependent thermal conductivity.
	
	The initial condition at time $t=0$ is given as: 
	\begin{equation*}
		T=T_0 \s \mathrm{in} \s \Omega,
		\centering
		\label{eq:initialCondition}
	\end{equation*}
	while the boundary conditions are: 
	\begin{align*}
		&T=\bar{T} \s\s\s \,\,\,\,\, \mathrm{on}  \s \partial \Omega_{T} \subset \partial \Omega,\\
		&k \nabla T \cdot \boldsymbol n  =  \bar{q}   \s\,\,  \mathrm{on}      \s  \partial \Omega_{Q} \subset \partial \Omega, \\
		&{k \nabla T \cdot \boldsymbol n  =  0   \s\,\,  \mathrm{on}      \s  \partial \Omega_{H} \subset \partial \Omega,}
	\end{align*}
	where $\bar{T}$ is the temperature of the environment on $\partial \Omega_{T}$,
	and $\bar{q}$ denotes the heat loss through the free surface of the normal $\boldsymbol n$ on $\partial \Omega_{Q}$.
	The quantity $\partial \Omega = \partial \Omega_T \cup \partial \Omega_Q \cup \partial \Omega_H$ is the boundary of the domain, $\Omega$,
	with $\partial \Omega_T$ the portion of $\partial \Omega$ where Dirichlet conditions are imposed,
	$\partial \Omega_Q$ the portion of $\partial \Omega$ where heat loss conditions are imposed,
	and $\partial \Omega_H$ the remaining portion of $\partial \Omega$ with adiabatic conditions.
	
	Since we use a part-scale PBF-LB/M thermomechanical model, the heat loss through the boundary $\partial \Omega_Q$
	can be described by means of a term $q_{p}$ representing the heat loss by conduction through
	the powder and a term $q_{g}$ representing the heat loss by convection through the environment gas. 
	Therefore, the heat loss $\bar{q}$ is given by:
	\begin{equation*}
		\bar{q}=q_{p}+q_{g}.
	\end{equation*}
	In a part-scale PBF-LB/M thermomechanical model, it is difficult to distinguish between the two heat transfer mechanisms \cite{chiumenti2017numerical}. 
	Therefore, explicit modelling of the unmelted powder in the surrounding environment is not required. Instead, the heat loss through the powder can be computed by a simplified approach using a convective boundary condition at the interface between powder and metal part. 
	Analogously, the heat loss through the environment gas can be computed using a convective boundary condition at the environment-gas-metal part interface.
	Consequently, both heat loss mechanisms can be modelled by two convective heat transfer coefficients: a powder convection coefficient, $h_p$, and a gas convection coefficient, $h_g$.
	Therefore, following Newton's law, we can express the two heat loss terms as follows: 
	\begin{equation*}
		\label{eq:hg}
		q_{p} = h_p(T-\bar{T}) \s\s \,\, \mathrm{on}  \s \Gamma_{p} \subset \partial \Omega_{Q},
	\end{equation*}
	\begin{equation*}
		q_{g} = h_g(T-\bar{T})  \s\s \,\, \mathrm{on}  \s \Gamma_{g} \subset \partial \Omega_{Q},
		\label{eq:hp}
	\end{equation*}
	where $\Gamma_{p}$ is the interface between the powder and the metal part and $\Gamma_{g}$ is the environment-gas-metal part interface, with $\partial\Omega_{Q} = \Gamma_{p} \cup \Gamma_{g}$ and $\Gamma_{p} \cap \Gamma_{g}=\emptyset$.
	
	\subsection{Mechanical problem}
	\label{subsec:Mc}
	To solve the PBF-LB/M mechanical problem, we use a quasi-static structural analysis where the mechanical response of the metal part is computed using a thermal load derived from solving the thermal problem. 
	The equilibrium equation governing the mechanical problem is given by:
	\begin{equation*}
		\nabla \cdot {{\boldsymbol \sigma}} = \boldsymbol 0 \s \mathrm{in} \s \Omega,
		\label{eq: eq10}
	\end{equation*}
	where $\boldsymbol{\sigma}$ is the second order Cauchy stress tensor defined as:
	\begin{equation*}
		{\boldsymbol \sigma} = {\boldsymbol D}^{el} {\boldsymbol \varepsilon}^{el},
		\label{eq: eq11}
	\end{equation*}
	with ${\boldsymbol D}^{el}$ the isotropic elasticity tensor, which depends on the Young's modulus of elasticity and Poisson's ratio, and ${\boldsymbol \varepsilon}^{el}$ is the elastic strain.
	The total strain in the material, ${\boldsymbol \varepsilon}^{tot}$, can be decomposed into the elastic strain, ${\boldsymbol \varepsilon}^{el}$, the thermal strain, ${\boldsymbol \varepsilon}^{th}$, and the plastic strain, ${\boldsymbol \varepsilon}^{pl}$, as follows:
	\begin{equation*}
		{\boldsymbol \varepsilon}^{tot} =  {\boldsymbol \varepsilon}^{el} + {\boldsymbol \varepsilon}^{th} + {\boldsymbol \varepsilon}^{pl} = \frac{1}{2} [\nabla \boldsymbol{u} + (\nabla \boldsymbol{u})^T], 
		\label{eq: eq12}
	\end{equation*}
	where $\boldsymbol u$ is the displacement vector.
	
	The thermal strain, which drives the mechanical problem and acts as an external thermal load, can be evaluated as:
	\begin{equation*}
		{\boldsymbol \varepsilon}^{th}  = \alpha\Delta T \boldsymbol I ,
		\label{eq:thermalStrain}
	\end{equation*}
	where $\alpha = \alpha(T)$ is the temperature dependent coefficient of thermal expansion, $\Delta T$ is the variation in time of the temperature field, and $\boldsymbol I$ is the second order identity tensor.
	
	Finally, the plastic strain rate $\dot{\boldsymbol \varepsilon}^{pl}$ is computed according to  the  Prandtl-Reuss flow rule \cite{chiumenti2017numerical,carraturo2020numerical,carraturo2021immersed} as follows:
	\begin{equation*}
		\dot{\boldsymbol \varepsilon}^{pl} = \dot{\gamma} \dfrac{\partial \Sigma}{\partial \boldsymbol \sigma}, 
		\label{eq: eq15}
	\end{equation*}
	where $\gamma$ is the equivalent plastic strain, $ \Sigma= \sigma_{\text{vm}}- \sigma_y \leq 0$ is the yield function describing the material through the equivalent Von Mises stress, $\sigma_{\text{vm}}=\sqrt{\frac{3}{2}\boldsymbol{s}:\boldsymbol{s}}$ with $\boldsymbol{s}=\boldsymbol{\sigma}-tr(\boldsymbol{\sigma})\boldsymbol{I}$, and the temperature dependent yield stress, $\sigma_{y}=\sigma_y(T)$ \cite{chiumenti2017numerical,carraturo2020numerical,carraturo2021immersed}.
	
	The mechanical problem is solved with the following Dirichlet boundary condition:
	\begin{equation*}
		\label{eq:u}
		\boldsymbol u=\boldsymbol 0 \s\s\s \,\,\, \mathrm{on}  \s \partial \Omega_{U} \subset \partial {\Omega},
	\end{equation*}
	where $\partial \Omega_{U}$ is the portion of the boundary $\partial \Omega$ where the fixed support is imposed; on the remaining part of the boundary $\partial \Omega$, we impose homogeneous Neumann boundary conditions.
	
	\section{Experimental setup}
	\label{sec:ExperimentalSetup}
\begin{figure}[!ht]
		\begin{center}
			\includegraphics[width=0.9\textwidth]{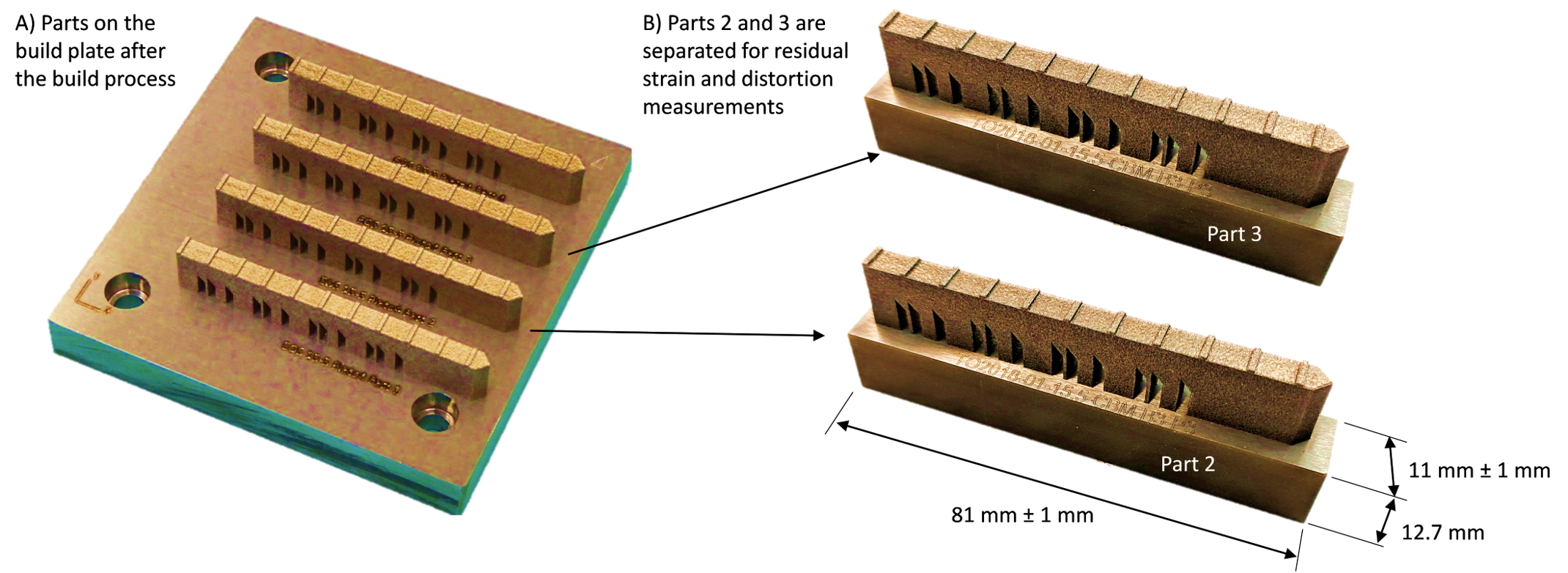}
		\end{center}
		\caption{AMBench2018-01 benchmark for the experiment performed by the NIST \cite{NIST,levine2020outcomes}.}
		\label{fig:geom}
	\end{figure}

	In the present section, we briefly describe the AMBench2018-01 benchmark for the experiment performed by NIST \cite{NIST,levine2020outcomes} which we referred to for our study.
	The geometry for the experiment is a bridge structure as shown in \Cref{fig:geom}. 
	All information related to the design and setup of the experiment reported here are available on the AMBench2018 website \cite{NIST}.
	As shown in \Cref{fig:geom}, four identical metal parts (cantilever beams) were constructed on the build platform for each construction process. 
	Only one cantilever beam is considered here because the four structures are spaced so that they do not interact with each other. 
	However, in order to take into account the cooling time of the layers, it is necessary to consider the entire build layout. 
	Therefore, the model considered in the present work, is a single cantilever beam, built on the build platform of 85 mm x 12 mm x 20 mm (length x height x width).
	
	As specified in \cite{NIST} the experiment is carried out using both 15-5 stainless steel and Inconel 625 super alloy. 
	In the present work, we consider only the latter \Lorenzo{alloy, whose} properties are discussed in \Cref{subsubsec:matmar}. 
	In addition, two build processes were proposed using two different machines: the Additive Manufacturing Meteorological Testbed (the NIST in-house build machine) and an EOS M270, which will be referred to herein. 
	In \Cref{tab:procmatA}, we report the process parameters used during the fabrication process with the EOS M270 machine, which were then used for our PBF-LB/M numerical simulations. 
	\begin{table}[!ht]
		\caption{Summary of the parameters of the EOS M270 machine setup used for the AMBench2018-01 benchmark of the NIST experiment \cite{NIST}.}
		\label{tab:procmatA}
		\begin{center}
			\begin{tabular}{l c}
				\hline
				\textbf{Parameters}  &   \textbf{Value} \\ \hline 
				Total number of  layers  &  ${ 625}$  \\
				Average layer time  &  ${52s}$  \\
				Layer height	 &  ${20\mu m}$  \\
				Contour scan speed  &   ${900 \frac{mm}{s}}$  \\
				Infill scan speed  & ${800 \frac{mm}{s}}$  \\
				Hatch space & ${100\mu m}$  \\
				\hline
			\end{tabular}
		\end{center}
	\end{table}
	After fabrication, part of the final structure is removed from the build platform by electrical discharge machining, allowing the part to flex upward due to residual stresses. The 11 ridges made on the top surface of the structure (see \Cref{fig:geom}) were ground to a smooth surface to accurately predict the displacement.
	
	Several measurements were made at the NIST laboratories, including the upward displacement of the beam after partial removal of the part from the construction platform. In the present work, we aim to replicate the residual strains and the upward displacement before and after removal of the metal part from the build platform, respectively.
	
	\section{Numerical simulations}
	\label{sec:numericalSimulations}
	
	In this section we discuss the numerical simulations performed by the Ansys2021-R2 software to reproduce the AMBench2018-01 benchmark described in \Cref{sec:ExperimentalSetup}. 
	
	The software uses a weakly coupled thermomechanical approach based on the finite element method \cite{chergui2021finite}, where the thermal and mechanical analyses are solved in a staggered manner for computational convenience.
	Specifically, the transient thermal analysis is solved first at each time step. 
	Then the thermal histories are used to perform the quasi-static mechanical analysis. \rev{All the numerical simulations of the PBF process discussed in this section are obtained on an HPC server equipped with a CPU with 128 AMD EPYC 7702@1.67 GHz cores and 376 GB RAM.}
	
	\subsection{Meshing strategy}
	\label{subsub:GM}
	
\begin{figure}[!ht]
		\begin{center}
			\subfigure[PARAMETRI-1][]{\label{fig:pdf4}\includegraphics[width=0.45\textwidth]{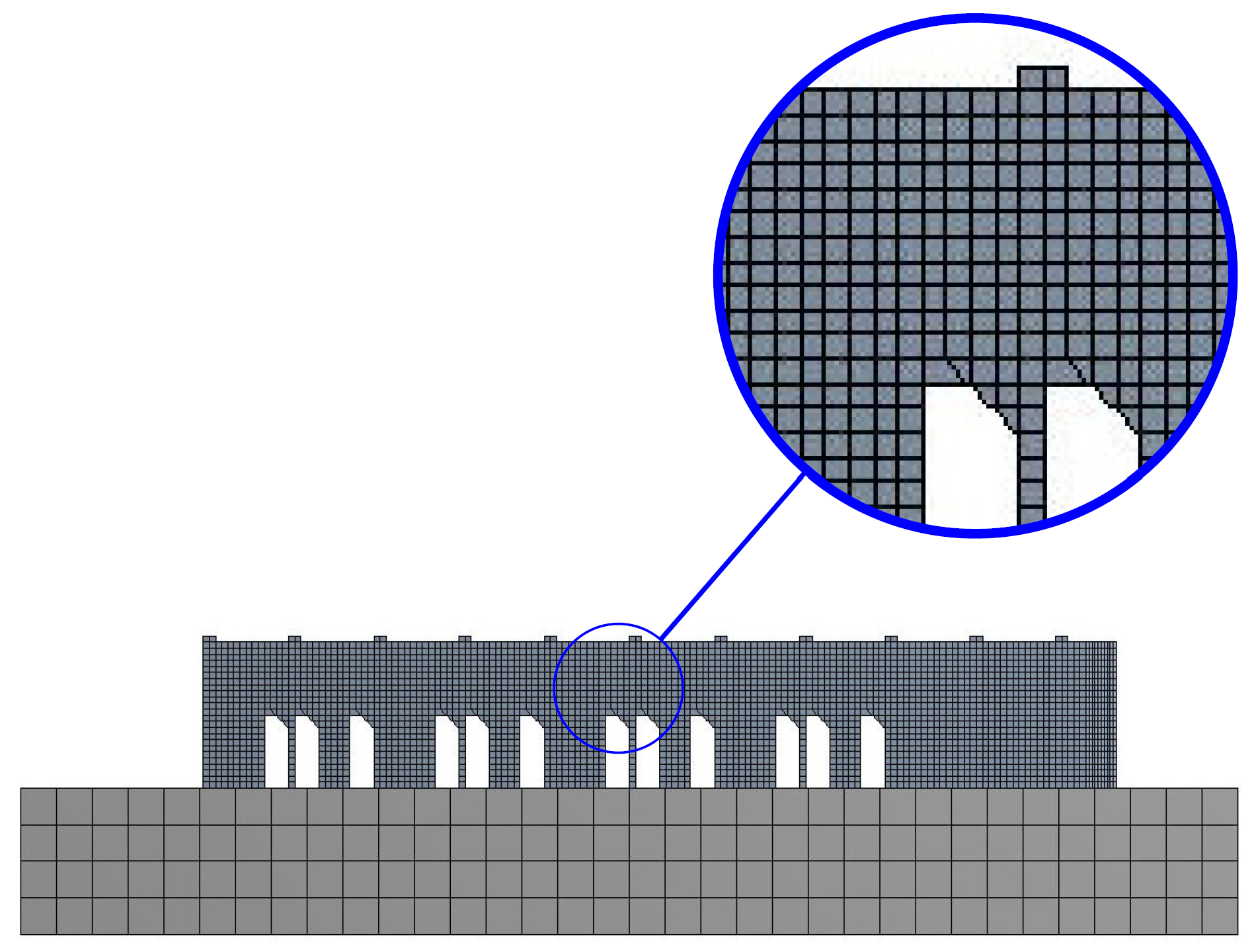}} \hspace{2em}
			\subfigure[PARAMETRI-2][]{\label{fig:pdf999}\includegraphics[width=0.45\textwidth]{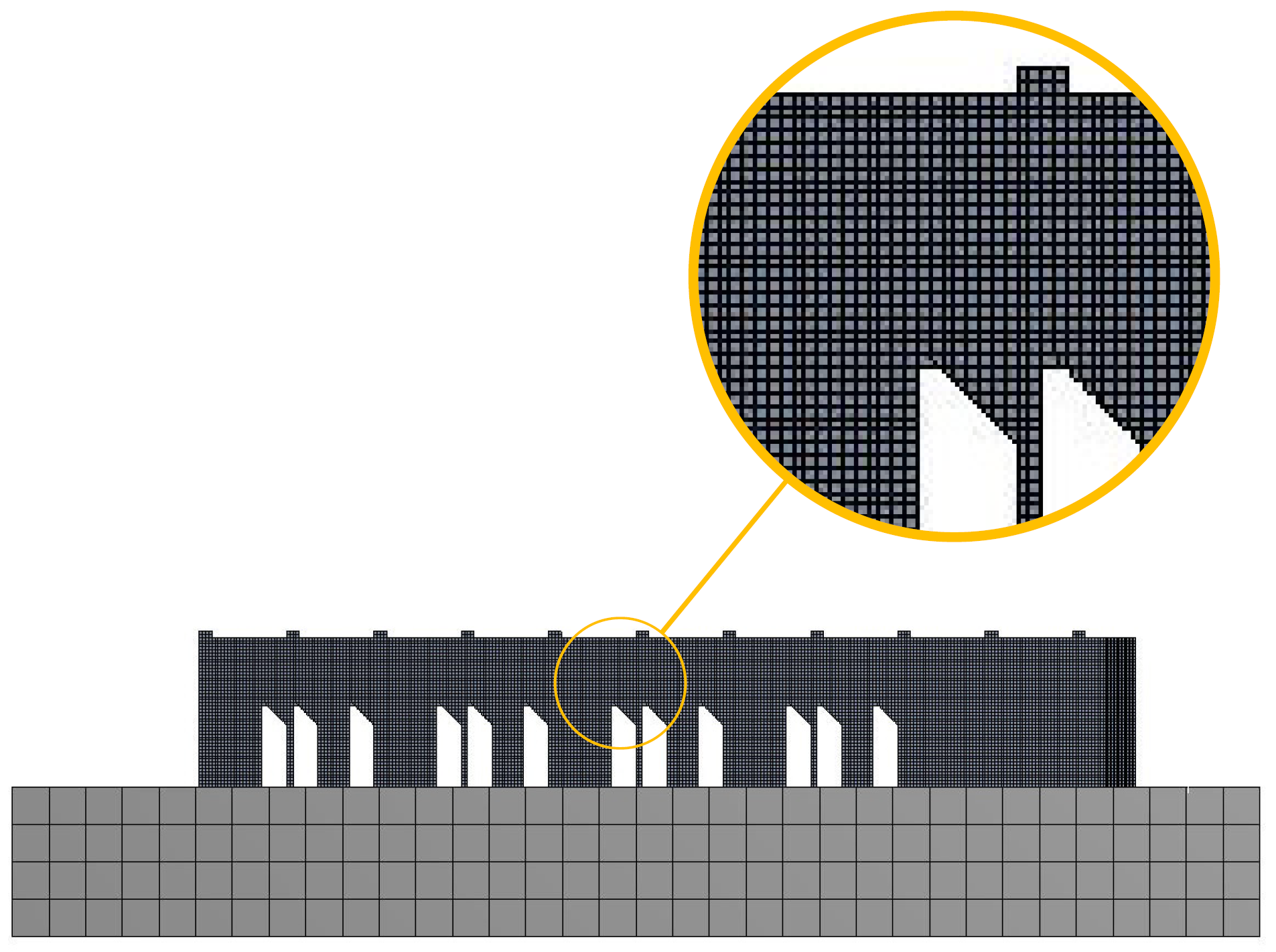}}
		\end{center}
		\caption{Meshing strategy for the part-scale PBF-LB/M thermomechanical model of the Inconel 625 cantilever beam: $(a)$ coarse mesh, $(b)$ fine mesh.}
		\label{fig:mesh} 
	\end{figure}

	In all the numerical simulations carried out in the present work, we adopt a uniform mesh of quadratic hexahedral finite elements.
	In particular, we consider two different mesh sizes for the build platform and the cantilever beam (i.e., the metal part), respectively.
	Furthermore, we consider two different mesh sizes for the metal part (\Cref{fig:mesh}), a coarser mesh with an element size of 0.5 mm (\Cref{fig:pdf4}) and a finer mesh with element size of 0.25 mm (\Cref{fig:pdf999}). 
	In the following, we will refer to them as coarse and fine mesh. 
	Additional information on the meshing strategy for the metal part, including the numerical simulation times required to complete the PBF-LB/M process, is given in \Cref{tab:FE}. It is noteworthy that the numerical simulation time required by the cantilever beam model with the fine mesh is significantly higher than the time required by the coarse mesh.
	For the build platform we choose a mesh with characteristic element of size 3 mm (\Cref{fig:mesh}), since the build platform is modelled only to account for heat loss through the build platform and as a mechanical constraint for the metal part. 
	\begin{table}[!ht]
		\caption{Meshing strategy for the part-scale PBF-LB/M thermomechanical model for the Inconel 625 cantilever beam.
			Summary of the number of nodes (nb. nodes), the total number of elements (nb. elements) in the domain and the number of layers in $z$-direction (nb. $z$-layers)
			in relation to the meshing strategy chosen for the metal part, including the numerical simulation times required (sim. times) to complete the PBF-LB/M process.}
		\label{tab:FE}
		\begin{center}
			\begin{tabular}{l c c c c c}
				\hline
				&\textbf{mesh strategy}  &   \textbf{nb. nodes} & \textbf{nb. elements} & \textbf{nb. $z$-layers} &\textbf{sim. times (approx.)} \\ \hline 
				& mesh coarse	 &  ${141095}$ & $30340$ & $25$ & 4 hours\\
				& mesh fine	 &  ${1041027}$ & $240560$ & $50$ & 6 days\\			
				\hline
			\end{tabular}
		\end{center}
	\end{table}
	
	\subsection{Material properties}
	\label{subsubsec:matmar}
	
\begin{figure}[!ht]
		\begin{center}
			\includegraphics[width=0.6\textwidth]{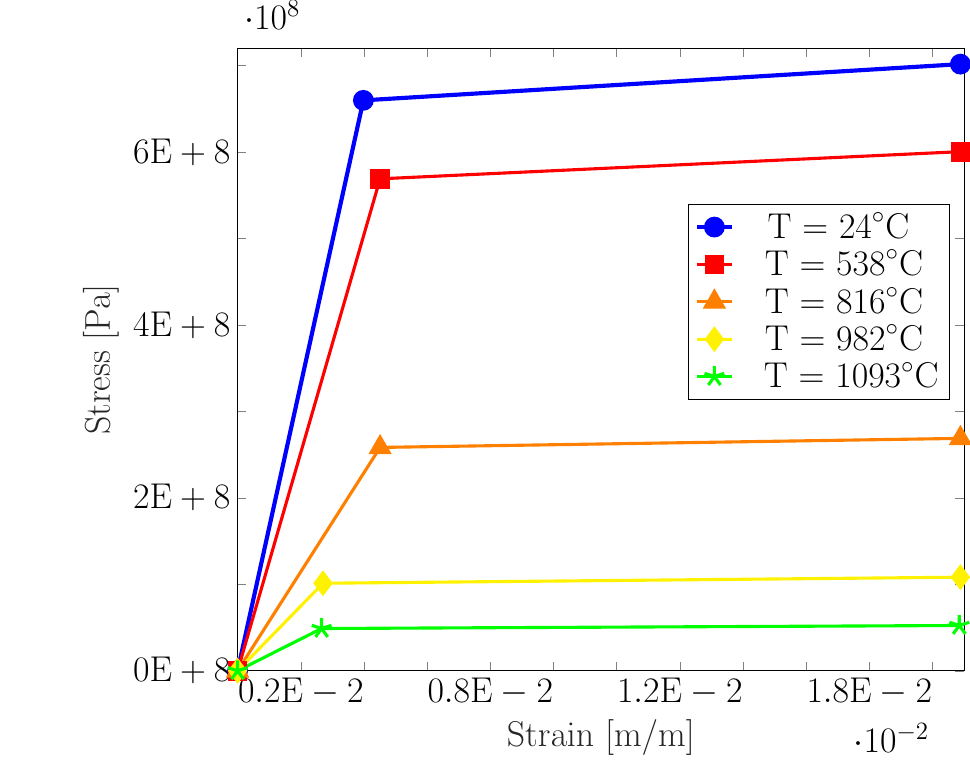}
		\end{center}
		\caption{Temperature dependent properties of Inconel 625 extrapolated from Ansys2021-R2:
			bilinear isotropic plastic hardening model with temperature dependent yielding behaviour.}
		\label{fig:Hardening}
	\end{figure}

	As specified in \Cref{sec:ExperimentalSetup}, the material used in the present work is the nickel based super alloy Inconel 625. 
	Although this alloy is widely used in industry and therefore its properties are well known, PBF-LB/M applications require specific material properties not only at room temperature, but also at temperatures close to melting \cite{Inconel}. 
	Therefore, there is the need for accurate measurements of both temperature dependent and temperature independent material properties.
	In the present work, all material properties are extrapolated from Ansys2021-R2 software.
	In particular, we set up a bilinear isotropic plastic hardening model with temperature dependent yielding behaviour, as shown in \Cref{fig:Hardening}. 
	All other temperature dependent material properties are shown in \Cref{tab:mat1,tab:mat2}. 
	Finally, we set the temperature independent material density and melting temperature to $8440$ ${\text{kg}}/{\text{m}}^3$ and $1290$ $^\circ{\text{C}}$, respectively.
	\begin{table}[!ht]
		\caption{Temperature dependent properties of Inconel 625 extrapolated from Ansys2021-R2: Young's Modulus and Poisson's Ratio.}
		\label{tab:mat1}
		\begin{center}
			\begin{tabular}{l c c c }
				\hline
				&\textbf{Temperature [$^\circ{\text{C}}$]} &\textbf{Young's Modulus [Pa] }  &\textbf{Poisson's Ratio [-] }  \\ \hline 
				&24	 &1.6E+11  &0.278 \\
				&538	 &1.2E+11  &0.305 \\
				&816	 &5.7E+11  &0.33 \\
				&982	 &3.8E+11  &0.33 \\
				&1093	 &2.1E+11  &0.33 \\
				\hline
			\end{tabular}
		\end{center}
	\end{table}
	\begin{table}[!ht]
		\caption{Temperature dependent properties of Inconel 625 extrapolated from Ansys2021-R2: coefficient of thermal expansion, specific heat, and thermal conductivity.}
		\label{tab:mat2}
		\begin{center}
			\begin{tabular}{l c c c c}
				\hline
				&\textbf{Temperature} &\textbf{Thermal expansion}  &\textbf{Specific Heat} &\textbf{Thermal Conductivity} \\
				&[$^\circ{\text{C}}$] &[$^\circ{\text{C}^{-1}}$] &[J $\text{kg}^{-1}$ $^\circ{\text{C}^{-1}}$] &[W $\text{m}^{-1}$ $^\circ{\text{C}^{-1}}$]\\ \hline 
				&93	&1.28E-05 &427 &10.8\\
				&204	&1.31E-05 &456 &12.5\\
				&316	&1.33E-05 &481 &14.1\\
				&427	&1.37E-05 &500 &15.7\\
				&538	&1.4E-05 &536 &17.5\\
				&649	&1.48E-05 &565 &19\\
				&760	&1.53E-05 &590 &20.8\\
				&871	&1.58E-05 &620 &22.8\\
				&927	&1.62E-05 &645 &25.2\\
				\hline
			\end{tabular}
		\end{center}
	\end{table}
	
	\subsection{Printing process}
	\label{subsec:amsimul}
	
	The numerical simulation of the PBF-LB/M process requires that the model evolves over time, i.e., elements are added using a layer-by-layer strategy. 
	We first construct the entire metal part with a layered mesh (\Cref{subsub:GM}) and then use the standard element birth and death technique to activate the element layers and simulate the progress of the construction. 
	More details can be found in \cite{chergui2021finite,michaleris2014modeling}.
	
	Another important aspect to consider is that simulating the entire construction process of a real part following the cantilever beam scanning pattern would require an enormous amount of computing time, making it impractical. 
	To achieve our goals in a reasonable amount of time -- i.e., much less than the actual build time -- we use the \lq\lq super-layer\rq\rq~abstraction.
	Specifically, the actual metal powder deposition layers are aggregated into finite element super-layers for simulation purposes. 
	Assuming that the thermal histories of the neighbouring physical layers are comparable, \rev{this approach is standard} and allows the simulation time to be contained. 
	Each super-layer is activated as a whole at the so-called activation temperature, $T_A$, and cools down for a period of time (i.e., dwell time) dependent on the process parameters (\Cref{tab:procma}) until the subsequent super-layer is activated. 
	In addition, to account for identical metal parts built at the same time (Section 3) but not present in our simulations, we set a dwell time multiplier of 4.
	
	To simulate the heat loss during the process, the build platform within the numerical model is assigned an initial temperature of $80$ $^\circ{\text{C}}$. 
	In addition, the construction part is assigned an initial temperature of $20$ $^\circ{\text{C}}$ since this is taken as the chamber temperature. 
	At the end of the process, all temperatures are raised to the chamber temperature. 
	The boundary condition in the quasi-static structural analysis is set by inserting a fixed support at the build  platform bottom surface (\Cref{eq:u}). 
	Finally, to be consistent with the AMBench2018-01 experiment \cite{NIST, levine2020outcomes}, we simulate a partial and progressive removal of the metal part from the build platform.
	\begin{table}[!ht]
		\caption{Summary of the machine setting parameters for the part-scale PBF-LB/M thermomechanical numerical simulation of the Inconel 625 cantilever beam model.}
		\label{tab:procma}
		\begin{center}
			\begin{tabular}{l c}
				\hline
				\textbf{Parameters}  &   \textbf{Value} \\ \hline 
				deposition thickness	 &  ${20\mu \text{m}}$  \\
				hatch space & ${100\mu \text{m}}$  \\
				number of heat sources & 1\\
				scan speed  &   ${900 \frac{\text{mm}}{\text{s}}}$  \\
				laser power &  ${100 \text{W}}$  \\
				\hline
			\end{tabular}
		\end{center}
	\end{table}
	
	\section{\Lorenzo{Uncertainty quantification workflow and surrogate modelling approach}}
	\label{sec:MFSG}
	
	In the present section, we discuss our proposed UQ approach. 
	Specifically, \Lorenzo{we introduce the UQ analyses that we wish to perform to quantify the uncertainties in the part-scale PBF-LB/M thermomechanical
		numerical model (thus reliably predicting the residual strains of the cantilever beam), and the mathematical framework needed to this end.
		In particular, we motivate the need for introducing a surrogate model of the PBF-LB/M numerical model, and describe the methodology
		that we consider to this end.}
	
	\subsection{Uncertainty quantification framework}
	\label{subsec:Bayes}
	
	In the present work, the interest is to quantify the uncertainties involved in the part-scale PBF-LB/M thermomechanical numerical simulations
	that depend on $N$ uncertain parameters. 
	
	We collect the uncertain parameters into a vector $\vv=(v_1,...,v_N)$ and denote by $\Gamma_n = [a_n, b_n]$
	the range of values for each uncertain parameter $v_n$, i.e., $\Gamma=\Gamma_1 \times ... \times \Gamma_N$ is the parameter space of $\vv$. 
	Finally, we denote the PDF of $\vv$ on $\Gamma$ by $\rho(\vv)$. 
	We further specify that the uncertain parameters considered are initially modelled as mutually independent uniform random variables
	in $\Gamma=\Gamma_{prior}$, i.e., we assume that they are described by $\rho_{prior}(\vv) = \prod_{n=1}^N \rho_{n,prior}(v_n)$ with $\rho_{n,prior}(v_n) = 1/(b_n - a_n)$.
	
	\Lorenzo{The first step of our UQ procedure consists in}
	a Bayesian inverse UQ analysis \cite{box2011bayesian,berger2013statistical,zhang2003compound,stuart:acta.bayesian,thanh-bui.gattas:MCMC}. 
	The goal of this analysis is to compute a data-informed PDF, also called posterior PDF, $\rho_{post}(\vv)$,
	in which a subset of parameter values is \lq\lq more likely" than others because it better matches the available data.
	In the present study, the available data are the displacements at the top of the cantilever beam model (\Cref{fig:geom}) based on the NIST experiment (\Cref{sec:ExperimentalSetup}).
	Using the data-informed posterior PDF derived by the Bayesian inverse UQ analysis, we \Lorenzo{then} perform a
	data-informed forward UQ analysis to \Lorenzo{reliably} predict residual strains in the middle plane of the cantilever beam. 
	
	\Lorenzo{The mathematical details of both steps are presented in \Cref{sec:resultsDiscussion}. However, we can already point
		that there is a critical issue in the UQ workflow outlined above. Indeed, several evaluations of the PBF-LB/M numerical model are required for both steps:
		during the inverse UQ, such evaluations are needed to \lq\lq test the compatibility" of the numerical predictions with the available data for different
		values of the uncertain parameters; during the forward UQ analysis instead, multiple evaluations are needed to assess the variability of the numerical
		prediction as the uncertain parameters vary according to their updated, data-informed, PDF.
		However, a single evaluation of the PBF-LB/M numerical model is considerably computationally expensive, especially on the finer of the two computational meshes that we consider in this work.  
		To overcome these issues, we introduce surrogate models of the PBF-LB/M numerical simulations, whose construction we briefly sketch in the next section.
		The aim of these surrogate models is to provide good approximations of the results of the expensive PBF-LB/M numerical simulations, at a much smaller computational cost.} 
	
	\subsection{Multi-fidelity MISC surrogate model}
	\label{subsec:MISCMISC}
	
	In the present work, we construct the surrogate models by the Multi-Index Stochastic Collocation (MISC) method, see
	\cite{hajiali.eal:MISC1,jakeman2019adaptive,piazzola.eal:ferry-paper,beck2019iga}, which is a multi-fidelity surrogate modelling technique
	that extends to the multi-fidelity paradigm the sparse grids method, one of the well-established single-fidelity surrogate modelling methodologies \cite{babuska.nobile.eal:stochastic2,xiu.hesthaven:high,piazzola2022sparse,b.griebel:acta}.
	
	To describe how MISC works, we denote by $\qoiscal(\vv)$ any Quantity of Interest (QoI) of the PBF-LB/M model, such as the displacements or the residual strains of the beam
	at a certain location. The values of these quantities can only be known approximately, by resorting to the numerical scheme introduced in \Cref{sec:numericalSimulations}, to solve the governing equations of the PBF-LB/M model on either of the two meshes detailed in \Cref{tab:FE}.
	We then introduce the notation $\qoiscal_\alpha(\vv)$, with $\alpha = 1, 2$ to refer to the value of any QoI computed
	on those two meshes, where $\alpha=1$ refers to the coarse mesh and $\alpha=2$ to the fine one. The approximations $\qoiscal_\alpha(\vv)$
	are called \emph{fidelities} in the following.
	Before continuing, we remark that the multi-fidelity approach described below can be extended easily to the case
	where more than two fidelities are available, as well as to the situation in which there is more than one discretisation parameter controlling
	the resolution of the numerical scheme (for instance, a mesh-size and a time-step). \rev{Furthermore, the different fidelities could also
	correspond to different descriptions of the physics of the problem: for instance, considering whether to simulate the physics of their printing process in a weakly coupled way (as detailed in the previous section) or in a fully coupled way (see, e.g., \cite{wang2021coupled}).} For a thorough discussion on multi-fidelity methods, see, e.g.,
	\cite{fernandez2016review,peherstorfer:mfsurvey}.
	
	The basic idea of the multi-fidelity surrogate modelling is to build an approximation of the function $\vv \mapsto \qoiscal(\vv)$
	given a set of $M$ evaluations of $\qoiscal_\alpha(\vv)$ obtained for suitable choices of $\vv$ and $\alpha$, where crucially
	most of the $M$ evaluations are obtained on the low-fidelity approximation corresponding to $\alpha=1$, hence keeping the computational cost
	necessary for the construction of the surrogate model to an acceptable level. Roughly speaking, the evaluations of $f_1(\vv)$
	are needed to capture the trend of the function $\vv \mapsto \qoiscal(\vv)$, whereas the few ones of $f_2(\vv)$ are only
	needed to correct the bias in the actual numerical values of the surrogate model introduced by having sampled the low-fidelity approximation only. 
	\rev{Therefore, it is evident that the success of the method largely depends on a sufficient correlation between the fidelities involved.
	The presence of a correlation between different fidelities is not always guaranteed and may be less common than sometimes believed: one situation in which correlation is present is when the different fidelities consists of different computational meshes, all of which sufficiently accurate to capture the essence of the physics (like in the current work).}
	In detail, the MISC multi-fidelity surrogate model of $\qoiscal(\vv)$ is computed as a linear combination of several single-fidelity surrogate
	models of $\vv \mapsto \qoiscal_\alpha(\vv)$
	each obtained by evaluating $\qoiscal_\alpha$ over a different set of values of $\vv$, according to the following equation:
	\begin{equation}
		\label{eq:surrogatoM}
		\qoiscal(\vv)  \approx \mcS_{\mcI} \qoiscal(\vv) = \sum_{[\alpha,\bbeta] \in \mcI} c_{\alpha,\bbeta} \mcU_{\alpha,m(\bbeta)}(\vv).
	\end{equation}
	In the equation above:
	\begin{enumerate}
		
		\item \rev{$m(\cdot)$ is a non-decreasing function; in this work we consider the following definition, but other choices are possible (see, e.g., \cite{piazzola2022sparse}):
		\begin{equation*}\label{eq:lev2knots}
			m:\mathbb{\mathbb{N}}_+ \rightarrow \mathbb{N}_+ \text{ such that } m(i) =2i-1.
		\end{equation*}}
		
		\item \label{item:beta} $\bbeta = (\beta_1, \ldots, \beta_N)\in \mathbb{N}^{N}_+$ is a multi-index, whose components control the sampling of the parameter space $\Gamma$.
		More specifically, we consider a Cartesian grid over $\Gamma$, $\mathcal{T}_{m(\bbeta)}$, composed by $m(\beta_1) \times m(\beta_2) \times \cdots \times m(\beta_N)$ points;
		we further denote by $\mathcal{T}_{n,m(\beta_n)}$ the $N$ sets of univariate points used to generate $\mathcal{T}_{m(\bbeta)}$.
		For efficiency reasons, the sets $\mathcal{T}_{n,m(\beta_n)}$ should be chosen according to the PDF $\rho_n$, and such that $\mathcal{T}_{n,m(r)} \subset \mathcal{T}_{n,m(s)}$
		if $r < s$, i.e., the collocation points must be nested.
		A good choice for $\rho_n$ uniform as in our problem is represented by the so-called symmetric Leja points \cite{narayan:leja,nobile.etal:leja}.
		The Cartesian grid $\mathcal{T}_{m(\bbeta)}$ is then the set of points
		$\mathcal{T}_{m(\bbeta)} = \left\{\vv \in \Gamma : v_n = v_{n,m(\bbeta)}^{(j_n)}: j_n \leq m({\beta}_n) \right\}$.
		We disclose already that after the Bayesian inversion the PDFs $\rho_n$ of the uncertain parameters will no longer be uniform,
		so that these points will need to change and the MISC surrogate model will need to be re-built. 
		
		\item $\mcU_{\alpha,m(\bbeta)}(\vv)$ is an $N$-variate Lagrangian interpolant of $\qoiscal_\alpha(\vv)$ built over the Cartesian grid $\mathcal{T}_{m(\bbeta)}$, i.e.,
		\begin{equation*}
			\label{eq:interp_tensor2}
			\qoiscal_\alpha(\vv) \approx \mcU_{\alpha,m(\bbeta)}(\vv) := \sum_{\jj \leq m({\bbeta})} f_\alpha\left(v_{m(\bbeta)}^{(\jj)}\right) \mathcal{L}_{m(\bbeta)}^{(\jj)}(\vv),
		\end{equation*}
		where $\jj \leq m(\bbeta)$ means $j_n \leq m(\beta_n)$ for $n = 1,\ldots,N$, and 
		$\mathcal{L}_{m(\bbeta)}^{(\jj)}$ are the $N$-variate Lagrange polynomials associated to the points in $\mathcal{T}_{m(\bbeta)}$.\footnote{i.e.,
			$\displaystyle \mathcal{L}_{m(\bbeta)}^{(\jj)}(\vv) = \prod_{n=1}^{N} \ell_{n,m(\beta_n)}^{(j_n)}(v_n)$ with 
			$\displaystyle \ell_{n,m(\beta_n)}^{(j_n)}(v_n) = \prod_{k=1, k\neq j_n}^{m(\beta_n)} \frac{v_n-v_{n,m(\beta_n)}^{(k)}}{v_{n,m(\beta_n)}^{(k)}-v_{n,m(\beta_n)}^{(j_n)}}$. }
		
		
		\item $\mcI \subset \mathbb{N}^{1+N}_+$ is a multi-index set, whose elements are extended multi-indices $[\alpha,\bbeta]$. It gathers all the $N$-variate Lagrangian interpolants that compose the MISC approximation.
		It is required that $\mcI$ is a downward closed set, i.e., such that
		\[
		[\gamma, \boldsymbol{\delta}] \leq  [\alpha,\bbeta] \mbox{ and } [\alpha,\bbeta] \in \mcI \Rightarrow [\gamma, \boldsymbol{\delta}] \in \mcI.
		\]
		The performance of the MISC surrogate model (i.e., its accuracy as the number of collocation points increase) crucially depends on the choice of the multi-index set $\mcI$.
		In this work, we consider an adaptive a-posteriori strategy to simultaneously enlarge the multi-index set $\mcI$ and refine the corresponding MISC surrogate model $\mcS_{\mcI} \qoiscal(\vv)$,
		see \cite{jakeman2019adaptive,piazzola.eal:ferry-paper} for details. Roughly speaking, at each iteration of such adaptive algorithm:
		\begin{enumerate}
			\item a few candidate extended multi-indices $[\alpha,\bbeta]$ are temporarily added to $\mcI$;
			\item the new corresponding MISC surrogate model $\mcS_{\mcI} \qoiscal(\vv)$ is computed,
			which requires solving the PBF-LB/M model equation on the mesh specified by $\alpha$ for the values of the parameters dictated by $\bbeta$;
			\item an heuristic selection criterion selects the extended multi-indices $[\alpha,\bbeta]$ that was most effective in improving the surrogate model \rev{\cite{jakeman2019adaptive,piazzola.eal:ferry-paper}};
			\item the list of candidates is updated accordingly, and the next iteration begins.
		\end{enumerate}
		
		\item $c_{\alpha,\bbeta}$ are the so-called combination technique coefficients defined as follows:
		\begin{equation*}
			\label{eq:coeff}
			\quad c_{\alpha,\bbeta}: = \sum_{\substack{i \in \{0,1\}, \jj \in \{0,1\}^{N}: \\ [\alpha+i,\bbeta+\jj] \in \mcI}} (-1)^{\lVert \jj \rVert_1}.
		\end{equation*}
		
	\end{enumerate}

	\section{Results of the UQ workflow}
	\label{sec:resultsDiscussion}
	
	In the present section, we show and discuss the numerical results obtained \Lorenzo{while performing the UQ workflow described in \Cref{sec:MFSG}.} 
	\Lorenzo{In details, we first present a procedure to compute the data-informed PDF of the uncertain parameters by Bayesian inversion techniques 
		and then illustrate the data-informed forward UQ analysis for the residual strains. In particular,
		we will highlight how this two-step procedure leads to an estimate of the uncertainty on the residual strains quite smaller than what we would have obtained
		considering the a-priori PDF for the uncertain parameters rather than the data-informed ones.}

	\subsection{Bayesian inverse UQ analysis}
	\label{subsec:bayesianinverse}
	
	\subsubsection{Uncertain parameters and available data}
	\label{subsubsec:uncertaindANDQoIs}
	
	In the present study
	we consider the powder convection coefficient, $h_p$ (see \Cref{subsec:Th}), and the activation temperature, $T_A$ (see \Cref{subsec:amsimul}), as uncertain parameters. 
	In our previous work \cite{chiappetta2022inverse}, we \Lorenzo{performed} a global sensitivity analysis to test the variability of the QoIs for different uncertain parameters,
	showing that $h_p$ and $T_A$ are the parameters that most influence \Lorenzo{the displacements of the beam}. 
	In addition, to ensure a minimally informed approach with respect to these parameters, we assume that each of them is treated as a random variable uniformly distributed over an appropriate interval,
	and that these two random variables are independent of each other; for the same reason, we take intervals (parameter space) larger than those typically used in the literature
	\cite{an2017neutron, gan2019benchmark, arisoy2019modeling, li2019numerical, hong2021comparative}, see \Cref{tab:PMat}. 
	\begin{table}[!ht]
		\caption{Parameter spaces of the uncertain parameters of the PBF-LB/M process simulation.}
		\label{tab:PMat}
		\begin{center}
			\begin{tabular}{ccc}
				\hline
				\textbf{uncertainty parameters}    & \textbf{units} & \textbf{parameter space} \\ \hline 
				powder convection coefficient ($\log$ $ h_{p}$) &  $- $  & $ [{-5};0]$\\
				activation temperature ($T_{A}$) &  $^\circ{C}$ & $[1130;1450]$  \\
				\hline
			\end{tabular}
		\end{center}
	\end{table}
	
	It is important to note that, although we could have considered $h_p$ directly as a uniform random variable, we choose to use its logarithm to increase the effectiveness of our approach. 
	This allows us to explore a wider range of values more easily and, at the same time, to give equal weight in the numerical analysis
	to the lowest and highest values within the range of $h_p$. 
	In other words, this choice allows us to investigate small values of $h_p$ with the same attention as larger ones.
	
\begin{figure}[!ht]
		\begin{center}
			\includegraphics[scale=0.45]{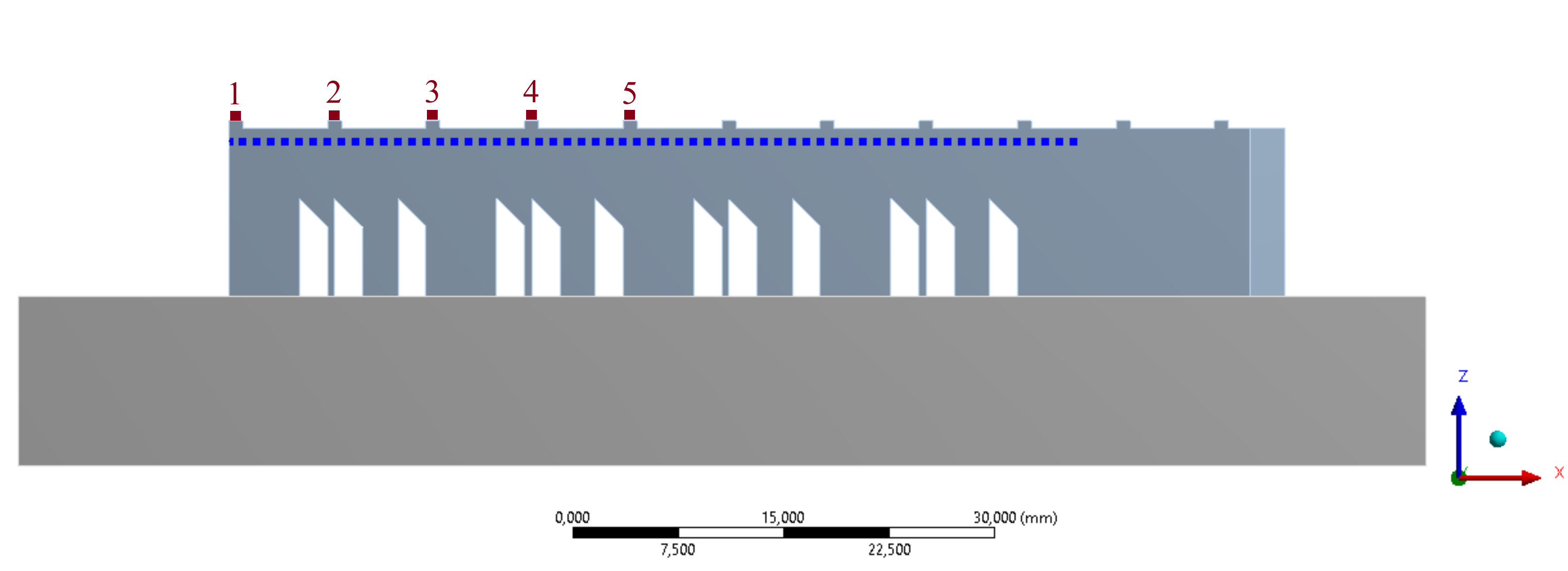}
		\end{center}
		\caption{PBF-LB/M numerical model of the cantilever beam 75 mm long, 12 mm high and 5 mm wide with a build platform measuring 85 mm long, 12 mm high and 20 mm wide.
			Points marked in purple are the {locations $\xx_{k,meas}, k=1,\ldots,5$} used during the Bayesian inverse UQ analysis;
			the (blue) dotted line marks the locations where we predict residual strains using data-informed forward UQ analysis.}
		\label{fig:dispstrain}
	\end{figure}

\begin{figure}[!ht]
		\begin{center}
			\includegraphics[width=0.5\textwidth]{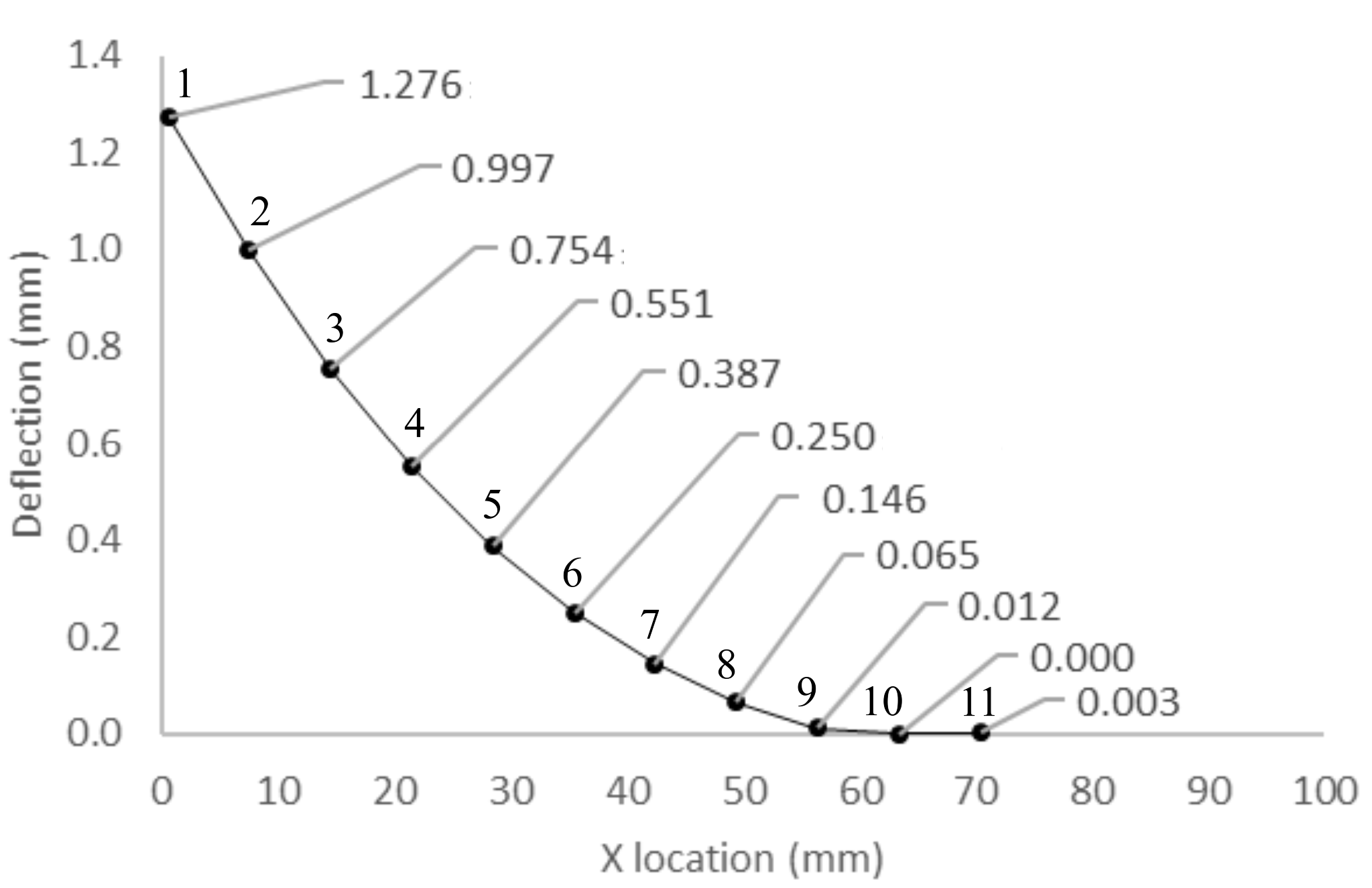}
		\end{center}
		\caption{Displacements at the 11 ridges at the top of the cantilever beam provided by the NIST \cite{NIST}.}
		\label{fig:NIST}
	\end{figure}

	\Lorenzo{The experimental data that we consider to perform the Bayesian inverse UQ analysis are the displacements at the first $K=5$ ridges
		of the cantilever beam provided by NIST \cite{NIST}, that we denote as $\xx_{k,meas}$, $k=1,\ldots,K$ (see \Cref{fig:dispstrain}).
		We decide to consider only the first 5 ridges of the beam as further tests with a larger number of data did not change the essence of the results shown below. 
		We assume that such experimental data \rev{correspond to} the solution of the PBF-LB/M model for an unknown value of the uncertain parameters $\vv_{true}$, corrupted by an experimental
		error, that is distributed as a Gaussian random variable with zero mean and standard deviation $\sigma_{meas}$;
		furthermore, the experimental errors at the $K$ locations are statistically independent. In mathematical terms, we can then write:
		\begin{equation}
			\label{eq:the-data}
			u_{k,exp} = u(\xx_{k,meas},\vv_{true}) + \epsilon_k, \quad \epsilon_k \sim \mathcal{N}(0,\sigma_{meas}^2), \quad k=1,\ldots,K.
		\end{equation}
		The values of $u_{k,exp}, k=1,\ldots,K$ are reported in \Cref{fig:NIST}, together with the displacements at the 6 additional ridges that we do not consider, as already motivated.}
	
	\subsubsection{Surrogate models for inverse UQ}
	\label{subsubsec:surrInv}

\begin{figure}[!ht]
		\begin{center}
			\subfigure[PARAMETRI-1][]{\label{fig:SparseTD}\includegraphics[width=0.405\textwidth]{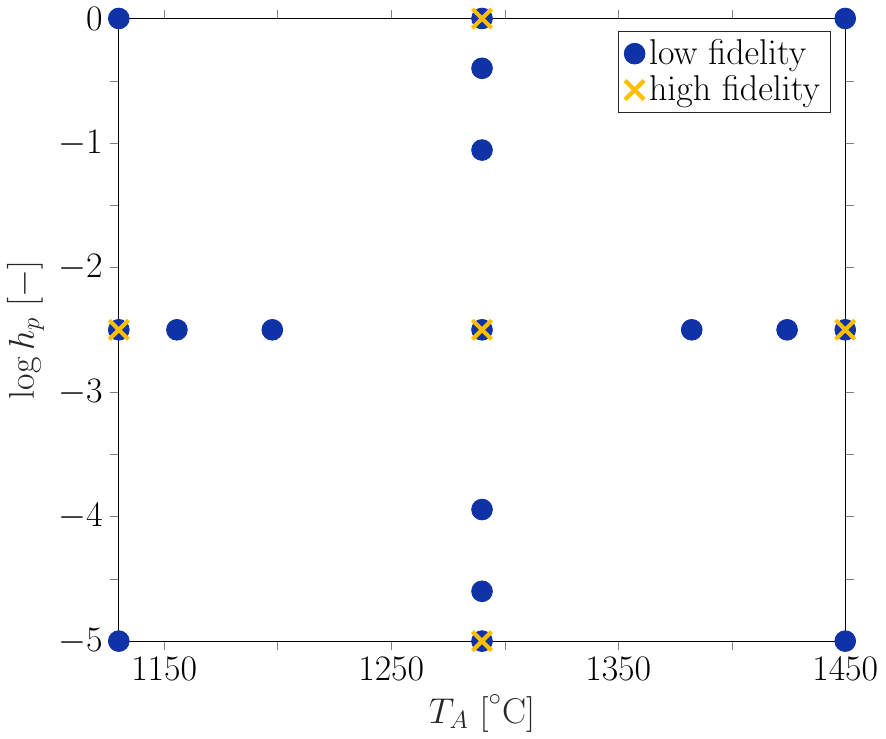}} \hspace{1.5em}
			\subfigure[PARAMETRI-2][]{\label{fig:x}\includegraphics[width=0.445\textwidth]{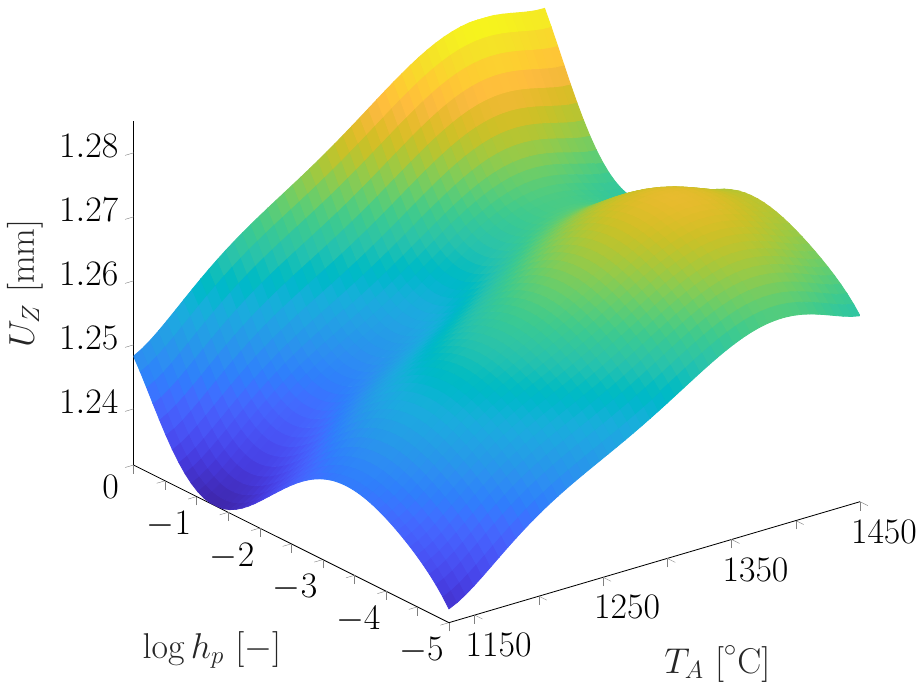}}
		\end{center}
		\caption{
			MISC surrogate model construction based on 17 evaluations (points marked in blue) of the low-fidelity PBF-LB/M numerical ($\alpha=1$)
			and 5 evaluations (points marked in yellow) of the high fidelity ($\alpha=2$) for the first node of the cantilever beam
			(i.e., $\xx_{1, meas}=(0.5, 2.5, 12.5)$): (a) sparse grid; (b) surrogate model.}
		\label{SparseGridtot_TD}
	\end{figure}

	\Lorenzo{The MISC surrogate model that we consider uses 22 evaluations of the PBF-LB/M numerical model,
		corresponding to the points in the parameter space represented in \Cref{fig:SparseTD}.
		In detail, we use 17 evaluations of the PBF-LB/M numerical model at the low fidelity, $\alpha=1$ and 5 at the high fidelity, $\alpha=2$.
		Note that the 5 points for which a high-fidelity simulation is requested are a subset of the 17 points for the low-fidelity simulation,
		due to the nestedness of the symmetric Leja points.  
	}  
	
	\Lorenzo{The surrogate models obtained for the $K$ locations $\xx_{k,meas}$ show similar behaviour;
		we plot in \Cref{fig:x} the one obtained for the first node of the cantilever beam, with coordinates
		$x=0.5$ mm, $y=2.5$ mm, $z=12.5$ mm.} 
	
	\rev{To further clarify our choice, it is important to remember that low fidelity is associated with the coarse mesh of the cantilever beam model, while high fidelity is associated with the fine mesh, as discussed in \Cref{subsub:GM}. When switching from low fidelity to high fidelity, we double the mesh size, i.e., the number of elements in the vertical direction of the cantilever beam model (\Cref{fig:mesh}), as shown in \Cref{tab:FE}. This choice is motivated by the desire to accurately capture the 11 ridges at the top of the beam in both models, as illustrated in  \Cref{fig:mesh}. The choice of adopting two levels of fidelity in our cantilever beam model is related to the fact that, although the low-fidelity model is able to represent the PBF-LB/M process, it is unable to capture all the details of the real geometry, in particular the openings in the cantilever beam  (see \Cref{fig:geom}). These details are captured by the high-fidelity model. Therefore, using the MISC methodology, we build the MISC model by exploiting the low-fidelity model to replicate the PBF-LB/M process and then correcting it with the high-fidelity model, which is crucial for more accurate results.}
	
\begin{figure}[!ht]
		\label{fig:Adap}
		\begin{center}
			\subfigure[PARAMETRI-1][]{\label{fig:comparison1}\includegraphics[width=0.3\textwidth]{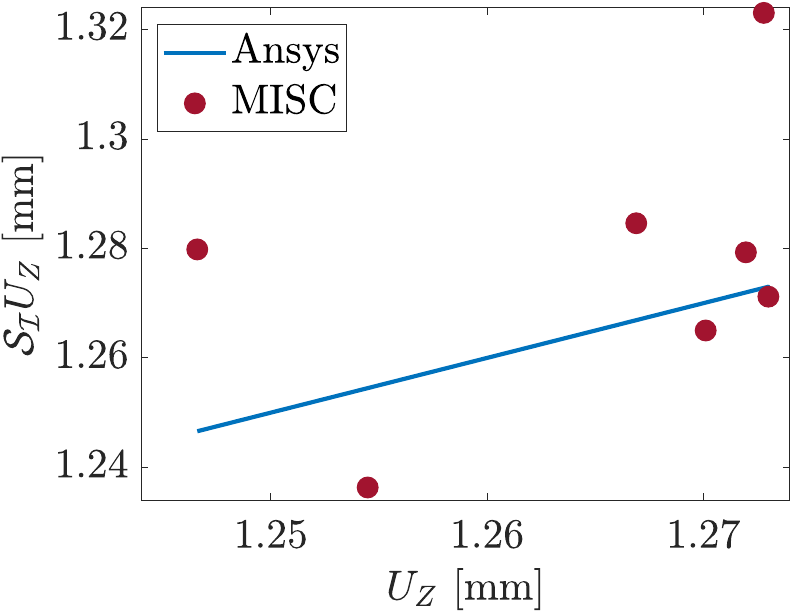}} 
			\subfigure[]{\label{fig:comparison2}\includegraphics[width=0.3\textwidth]{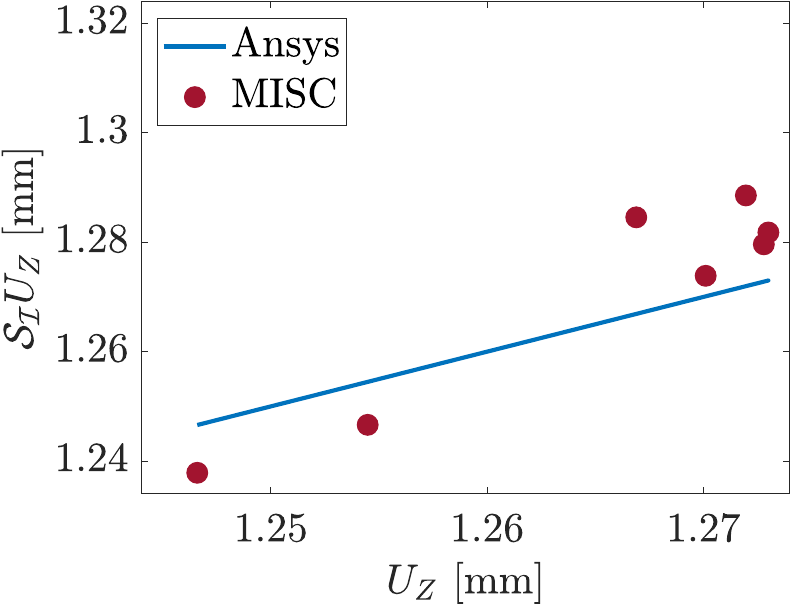}} 
			\subfigure[]{\label{fig:comparison3}\includegraphics[width=0.3\textwidth]{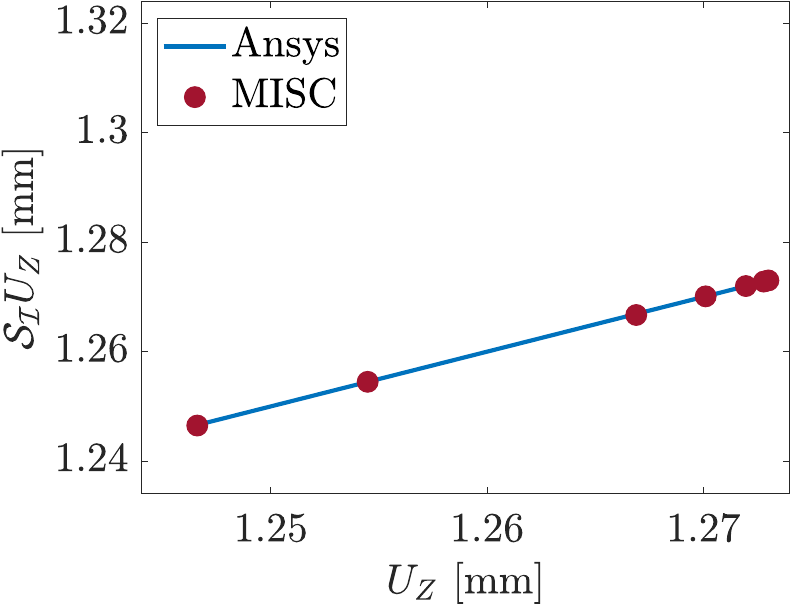}} 
		\end{center}
		{\caption{\rev{Results of the MISC surrogate model for inverse UQ analysis.
		Comparison between high-fidelity displacement results with ANSYS2021-R2, $U_{Z}$, and the displacement results for three different MISC surrogate models, ($\mathcal{S_I}U_{Z}$): a) MISC model with 6 (5+1) evaluations; b) MISC model with 10 (7+3) evaluations; c) MISC model with 22 (17+5) evaluations. The blue line represents the pairs ($U_{Z,ansys}, U_{Z,ansys}$) and the burgundy points the pairs ($U_{Z,ansys}, U_{Z,MISC}$), evaluated for each of the 7 validation points.
		}}}
	\end{figure}

	\rev{The adoption of the MISC surrogate model with 17 evaluations on the low-fidelity model and 5 evaluations on the high-fidelity model is driven by a performance analysis of the MISC surrogate model. We run the adaptive MISC algorithm detailed in \Cref{subsec:MISCMISC} until we obtained a surrogate model that provided acceptable pointwise prediction error and mean square error values over a specified set of validation points. This methodology allowed us to ensure that the surrogate model is reliable and accurate.
	To evaluate the effectiveness of the obtained MISC surrogate model, we compare the displacement results obtained through numerical simulations with ANSYS2021-R2 software on the high-fidelity model, $U_Z$, with the results interpolated from the MISC model, $\mathcal {S_I}U_Z$, for 7 couples of uncertain parameters $(T_A, h_p)$ values (validation points) represented by the burgundy points in Figure 7. In particular, in Figure 7 we show the comparison of the displacement results using: the MISC model based on 6 evaluations (5 on low-fidelity model and 1 on high-fidelity model), \Cref{fig:comparison1}, the MISC model based on 10 evaluations (7 on low-fidelity model and 3 on high-fidelity model), \Cref{fig:comparison2}, the MISC model based on 22 evaluations (17 on low-fidelity model and 5 on high-fidelity model), \Cref{fig:comparison3}. The analysis of the results revealed a remarkable degree of accuracy in the selected MISC surrogate model (\Cref{fig:comparison3}). In fact, contrary to what is shown in \Cref{fig:comparison1,fig:comparison2}, in \Cref{fig:comparison3} we can see that all points interpolated on the MISC model show significant alignment with the results obtained from ANSYS2021-R2 (represented by the blue line in Figure 7), indicating a good predictive ability of the surrogate model. This underlines the importance of an adequate number of evaluations to ensure the robustness and reliability of the MISC surrogate model.\\
		This level of consistency between the results obtained from the direct simulations and those interpolated from the MISC surrogate model suggests that the surrogate model is capable of effectively capturing the complex relationships between the input parameters and the responses of the thermomechanical beam model. Interestingly, this accuracy was achieved despite the MISC model being trained using a relatively small number of numerical evaluations, thus demonstrating the effectiveness of the MISC method in optimising the balance between accuracy and computational cost.}
	
	\subsubsection{Data-informed posterior PDF}
	\label{subsubsec:datainformed}
	
	The goal of Bayesian inverse UQ analysis is to find the \Lorenzo{so-called} posterior PDF of $\log h_p$ and $T_A$, indicated \rev{by} $\rho_{post}(\vv)$, i.e.,
	\Lorenzo{a PDF that assigns larger probability to regions of the parameter space that produce outputs that are more compatible with the available
		experimental data. To this end, we postulate that $\rho_{post}(\vv)$ can be well approximated by a Gaussian distribution,
		\rev{which is a common approach in literature}, see, e.g., \cite{kim.eal:hippylib,thanh-bui.gattas:MCMC,stuart:acta.bayesian,piazzola.eal:SIR}.}
	This approach \rev{requires setting the center} of the Gaussian approximation of $\rho_{post}(\vv)$ at 
	the so-called Maximum A Posterior (MAP) point of the true posterior PDF, $\vv_{MAP}$, i.e., 
	\begin{equation}
		\label{eq:ymapgg}
		\mathbf v_{MAP} :=  \argmax_{\vv \in {\mathbb{R}^N}} \rho_{post}(\vv),
	\end{equation}
	which can be roughly thought of as the most likely value of $\vv$ according to $\vv_{MAP}$.
	This quantity is actually computable, since the \rev{Bayes' theorem} allows us to derive an explicit expression for the posterior PDF:
	
	\begin{equation*}
		\label{eq:rhopost}
		\rho_{post}(\mathbf v) = \mathcal{L}(\vv \,|\, u_{1,exp}, u_{2,exp},\cdots) \ \rho_{prior}(\vv) \ \frac{1}{C}.
	\end{equation*}
	\Lorenzo{In the equation above, $C$ is a normalisation constant that makes $\rho_{post}(\mathbf v)$ actually a PDF (i.e., its integral equal to 1)
		and $\mathcal{L}(\vv \,|\, u_{1,exp}, u_{2,exp},\cdots)$ is the so-called likelihood function,
		which quantifies the plausibility that the measured displacements were generated by $\vv$.
		More specifically, under the assumption made in \Cref{subsubsec:uncertaindANDQoIs} that the experimental errors $\epsilon_k$ are statistically independent random variables with
		PDF $\rho_{\epsilon_k}$, cf. Equation \eqref{eq:the-data}, the likelihood function can be written as
		\[
		\mathcal{L}(\vv \,|\, u_{1,exp}, u_{2,exp},\cdots) = \prod_{k=1}^5 \rho_{\epsilon_k} \left( u_{k,exp} - u(\xx_{k,meas},\vv) \right).
		\]
	}
	
	\Lorenzo{Given the expression above for the likelihood, and under the further assumptions that
		the prior PDF of the uncertain parameters are \rev{independently  uniform}
		(i.e., $\rho_{prior}(\vv)$ is a constant over $\Gamma$ and zero elsewhere, cf. \Cref{subsec:Bayes}),
		and that the experimental errors are Gaussian random variables, cf. \Cref{eq:the-data},
		it can be easily shown \cite{kim.eal:hippylib,thanh-bui.gattas:MCMC,stuart:acta.bayesian,piazzola.eal:SIR}
		that the maximisation problem in Equation \eqref{eq:ymapgg}} is in practice equivalent to the classical least-squares approach for the calibration of $\vv$: 
	\begin{align} \label{eq:ymap}
		\vv_{MAP}
		& =  \argmin_{\vv\in {\Gamma}} [-\log \mathcal{L}(\vv \,|\, u_{1,exp}, u_{2,exp},\cdots)]
		=  \argmin_{\mathbf v\in {\Gamma}} {\sum_{k=1}^K \left( u_{k,exp} - u(\xx_{k,meas},\vv) \right)^2} \\
		& \approx  \argmin_{\vv\in    {\Gamma}} {\sum_{k=1}^K \left( u_{k,exp} - \mathcal{S}_\mcI u(\xx_{k,meas},\vv) \right)^2,}  \nonumber
	\end{align}
	\Lorenzo{where crucially in the last step we have replaced the full PBF-LB/M model evaluations $u(\xx_{k,meas},\vv)$
		with their MISC surrogate model approximations to speed up the computations.} 
	{Also, note that the minimisation in
		Equation \eqref{eq:ymap} does not depend on the standard deviation of the experimental errors, $\sigma_{meas}$.
		If needed, a sample estimate of this quantity can be obtained as
		\[
		\sigma_{meas}^2 \approx \frac{1}{K} \sum_{k=1}^K \left( u_{k,exp} - \mathcal{S}_\mcI u(\xx_{k,meas},\vv_{MAP}) \right)^2.
		\]
	}
	
	Once $\vv_{MAP}$ is obtained, we are left with computing the covariance matrix $\Sigma_{post}$
	of the Gaussian approximation of $\rho_{post}(\vv)$.
	Under the same assumptions that lead to \Cref{eq:ymap}, 
	it can be shown that
	$\Sigma_{post}$ is equivalent to the inverse of the Hessian of the least square functional in Equation \Cref{eq:ymap} \cite{kim.eal:hippylib,thanh-bui.gattas:MCMC,stuart:acta.bayesian,piazzola.eal:SIR}.
	A good approximation can be then computed as
	\[
	\Sigma_{post} \approx \frac{1}{\sigma_{meas}^2} \mathcal{J}_u^T\mathcal{J}_u,
	\]
	where $\mathcal{J}_u$ is the Jacobian of the PBF-LB/M model prediction $u(\xx_{k,meas},\vv)$ evaluated the the MAP point, i.e.,
	\[
	[\mathcal{J}_u]_{k,j} = \frac{\partial}{\partial v_j} u(\xx_{k,meas},\vv_{MAP}).
	\]
	\rev{In practice, the partial derivatives above can be computed, e.g.,\ by finite differences. Furthermore, it is useful to replace the full PBF-LB/M model evaluations by their MISC surrogate model approximations.}

	\Lorenzo{Following the steps just described, we obtain $\vv_{MAP}=(1386^\circ\text{C}; -3)$ and that $\Sigma_{post}$ is essentially diagonal,
		which means that $T_A$ and $\log h_p$ are essentially statistically independent, i.e., two independent Gaussian variables; the diagonal entries of   $\Sigma_{post}$ provide their standard deviations, $\sigma_{T_A,post}=95 \,{^\circ{\text{C}}}$ and $\sigma_{\log h_p,post}=0.92$.}
	It is important to note that for the calibration of the parameter $\vv$ (\Cref{eq:ymap}) we used the Nelder-Mead gradient-free optimisation algorithm, accessible in MATLAB through the \texttt{fminsearch} command. To increase the robustness of our results, we performed several optimisations with 20 different initial conditions for the optimisation algorithm requiring overall 10000 surrogate model evaluations.
	
	\begin{figure}[!ht]
		\begin{center}
			\subfigure[PARAMETRI-1][]{\label{fig:hppost1}\includegraphics[width=0.46\textwidth]{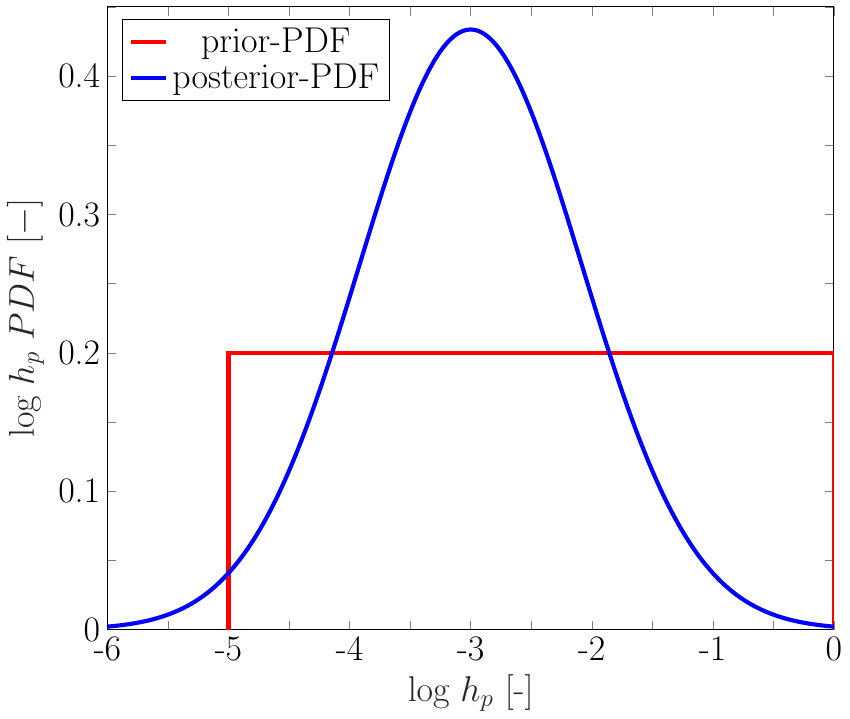}} 
			\subfigure[PARAMETRI-2][]{\label{fig:tapost1}\includegraphics[width=0.45\textwidth]{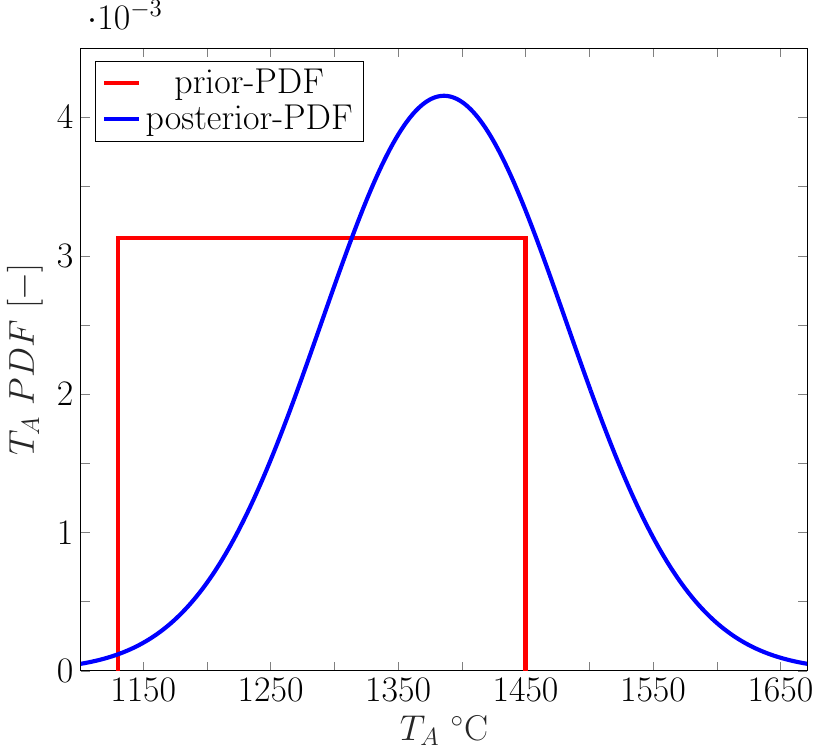}} 
		\end{center}
		\caption{Results of the Bayesian inverse UQ analysis: (a) uniform prior PDF and Gaussian posterior PDF for parameter $ \log h_p$ parameter before and after Bayesian inverse UQ analysis; (b) uniform prior PDF and Gaussian posterior PDF for parameter $T_A$ before and after Bayesian inverse UQ analysis.}
		\label{fig:beta}
	\end{figure}

	\begin{table}[!ht]
		\caption{Statistical information for the two uncertain parameters before and after the Bayesian inversion.}
		\label{tab:variance-red}
		\begin{center}
			\begin{tabular}{ccccc}
				\hline
				\textbf{Prior}	& \textbf{mean}		& \textbf{st. dev.}	& \textbf{CoV} 	& \textbf{interval} \\
				\hline 
				$T_A$		& $1290^\circ\text{C}$	& $92.37^\circ\text{C}$	& $0.072$	& $[1130; 1450]^\circ\text{C}$ 	\\
				$\log h_p$ 		& $-2.5$		& $1.44$		& $0.577$	&  $[-5; 0]$ 		 \\[15pt]
				\hline
				\textbf{Posterior}	& \textbf{mean}		& \textbf{st. dev.} 	& \textbf{CoV} 	& \textbf{``interval''} \\
				\hline 
				$T_A$		& $1386^\circ\text{C}$	& $95^\circ\text{C}$	& $0.068$	& $[1100; 1671]^\circ\text{C}$ 	\\ 
				$\log h_p$ 		&  $-3$			& $0.92$		& $0.306$	& $[-5.76; -0.24]$  		 \\
			\end{tabular}
		\end{center}
	\end{table} 
	
	In \Cref{fig:beta} we compare the prior and posterior PDFs of the two parameters.
	\Lorenzo{Note, in particular, that the posterior PDFs of both parameters significantly overflow from  the prior parameter space $\Gamma$. 
		More precisely, this means that there is a quite large set of values of the parameters outside of $\Gamma$ that are reasonably
		compatible with the experimental data, especially for $T_A$. Note however that this does not mean that we have not reduced the uncertainty
		on the values of the uncertain parameters compared to our a-priori assumptions on them. This can be checked upon comparing the
		standard deviations of the prior and posterior PDF of the parameters%
		\footnote{the standard deviation of a uniform random variable in $[a,b]$ is $\sqrt{\frac{(b-a)^2}{12}}$.},
		as well as their prior and posterior coefficients of variations (CoV), i.e.,
		the ratios between standard deviation and the absolute value of the mean of the uncertain parameters.
		Results are reported in \Cref{tab:variance-red}, and show that we are significantly reducing the uncertainty on $\log h_p$
		and slightly so for $T_A$. \Cref{tab:variance-red} also reports a rough comparison of the ranges for $T_A$ and $\log h_p$
		before and after the inversion procedure, based on the fact that a Gaussian random variable can roughly be considered to take
		values in the interval of length 6 standard deviations centered at its mean.
		\footnote{\rev{If $X$ is a Gaussian random variable with $\mu$ and standard deviation $\sigma$, a straightforward computation allows to determine the probability that $X$ takes values in said interval, as: \begin{equation}
					p(\mu-3 \sigma \leq X \leq \mu + 3 \sigma) = erf(3/\sqrt 2) \approx 0.9973.
		\end{equation}}}
	}

	\subsection{Data-informed forward UQ analysis}
	\label{subsec:datainfforward}
	
	
	The aim of the data-informed forward UQ analysis is to predict the profile of the residual strains of the cantilever beam and their associated uncertainties.
	Specifically, the quantities of interest are the residual strains of $J=120$ nodes located in the middle plane of the cantilever beam along the x-direction
	with a \Lorenzo{sampling step} 
	of $0.5$ mm and for $y=2.5$ mm and $z=11$ mm, ${\varepsilon}_{xx}(\xx_{j,str},\vv)$ with $j=1,...,J$ (\Cref{fig:dispstrain}). 
	
	\Lorenzo{To do this, we generate $S=10000$ samples of $\vv$ based on $\rho_{post}$ and for each one we evaluate ${{\varepsilon}_{xx}(\xx_{j,str},\vv)}$,
		which again requires introducing suitable MISC surrogate models. 
		We then estimate the PDF of the residual strains at each cantilever beam node by applying a kernel density estimation method \cite{parzen:kde,rosenblatt:kde}
		to the $S$ samples. Finally, we predict the residual strains of the cantilever beam by evaluating the mode of these PDFs.
		We discuss below the results obtained in this way, and refer the reader to \cite{chiappetta2022inverse} for more details on the methodology.}

	\subsubsection{Surrogate models for forward UQ analysis}
	\label{subsubsec:surrForw}
	
	\Lorenzo{Since the QoIs are now the residual strains of the beam, we need to build a new set of surrogate models, again by the MISC method.
		Note that the PBF-LB/M simulations that were run at the collocation points in \Cref{fig:SparseTD} for building the surrogate models
		for the Bayesian inversion are no longer usable, and we need to run a new set of simulations. This is for two reasons:
		first, those collocation points were selected based on the assumption that uncertain parameters were uniform random variables,
		whereas we are now using Gaussian random variables, which calls for a different set of collocation points to maintain good approximation
		properties (cf. \Cref{item:beta} of the enumerated list in \Cref{subsec:MISCMISC}). Second, even if we would be
		willing to settle for a suboptimal surrogate model, the surrogate model used for the Bayesian inversion is by construction only valid in $\Gamma$,
		whereas the support of $\rho_{post}$ significantly overflows such domain.}

\begin{figure}[!ht]
		\begin{center}
			\includegraphics[width=0.5\textwidth]{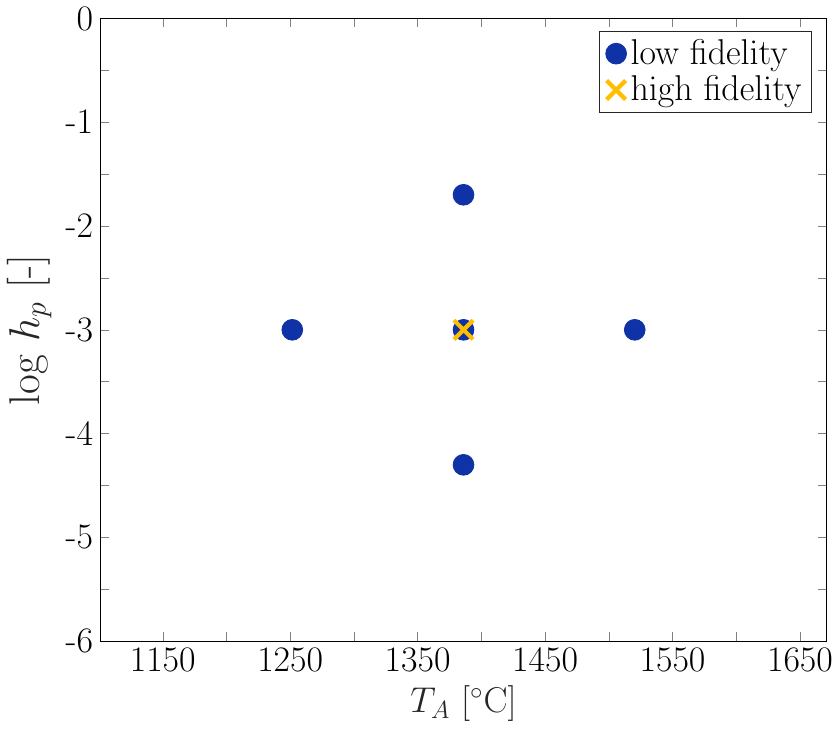}
		\end{center}
		\caption{Collocation points used for the construction of the MISC multi-fidelity surrogate models
			based on 5 evaluations (points marked in blue) of the PBF-LB/M numerical model at low fidelity, $\alpha=1$, and 1 evaluations (marked in yellow)
			of the PBF-LB/M numerical model at high fidelity, $\alpha=2$ for the residual strains cantilever beam model.}
		\label{fig:sparsefor}
	\end{figure}

\begin{figure}[!ht]
		\begin{center}
			\subfigure[PARAMETRI-1][$x = 0$ $\text{mm}$ ($J=1$)]{\label{fig:pdf5}\includegraphics[width=0.25\textwidth]{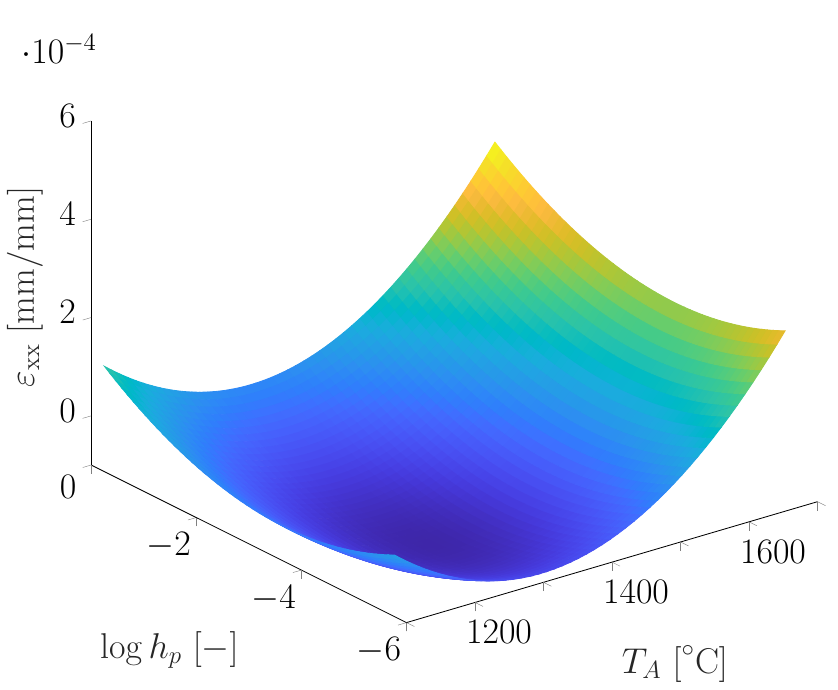}} 
			\subfigure[PARAMETRI-2][$x = 1.5$ $\text{mm}$ ($J=4$)]{\label{fig:pdf10}\includegraphics[width=0.25\textwidth]{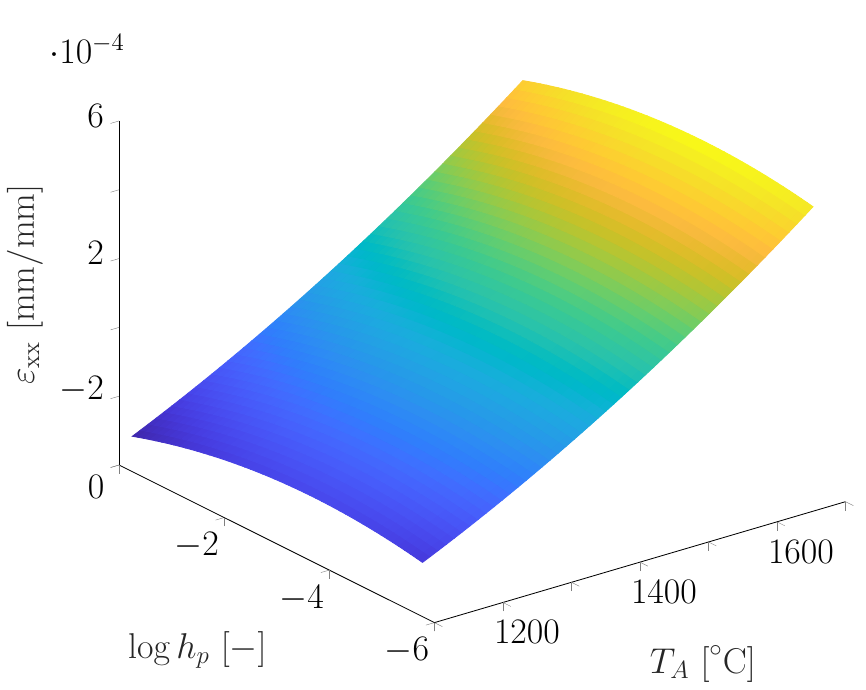}} 
			\subfigure[PARAMETRI-3][$x = 2.5$ $\text{mm}$ ($J=6$)]{\label{fig:pdf15}\includegraphics[width=0.25\textwidth]{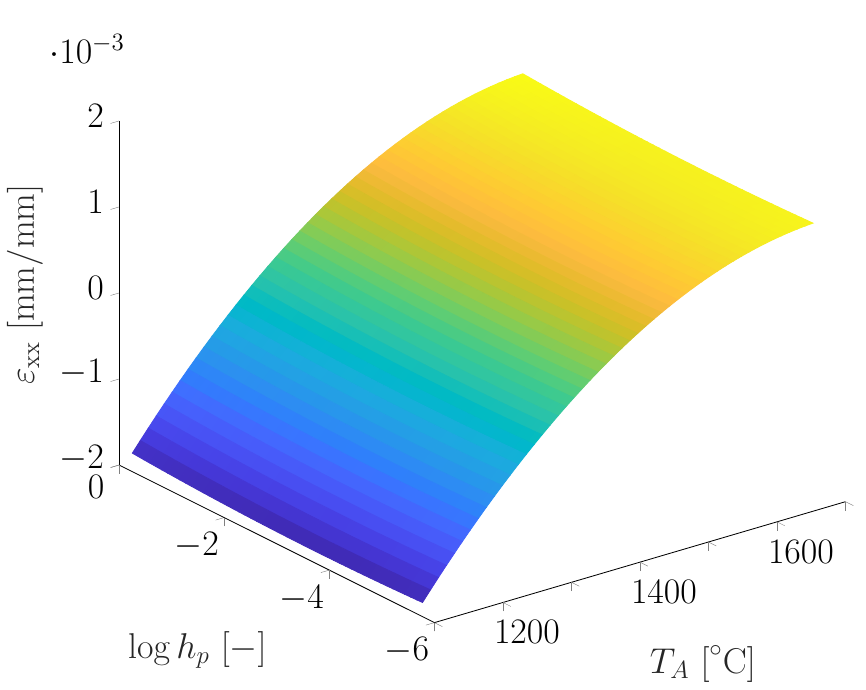}} 
			\subfigure[PARAMETRI-4][$x = 20.5$ $\text{mm}$ ($J=42$)]{\label{fig:pdf20}\includegraphics[width=0.25\textwidth]{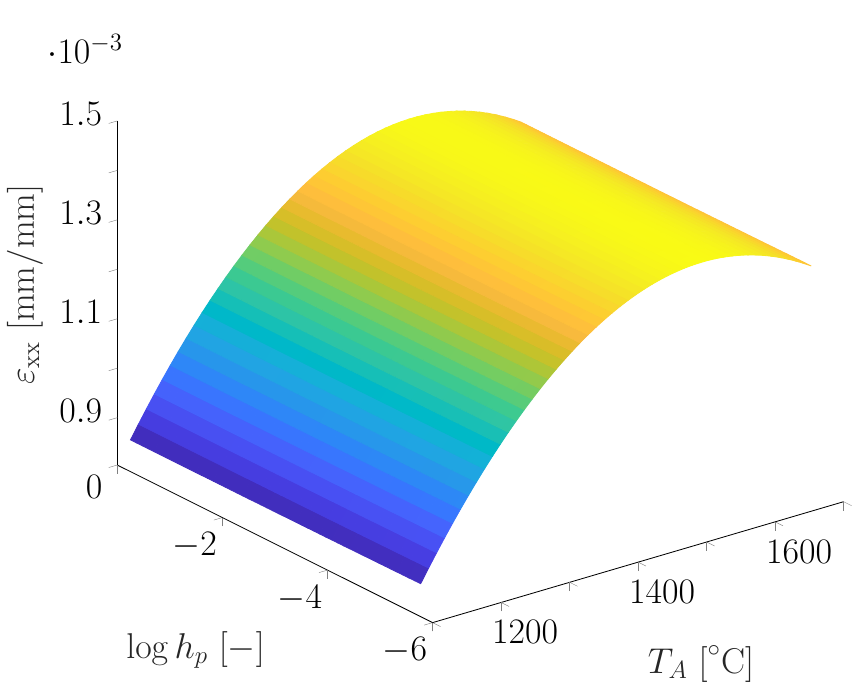}} 
			\subfigure[PARAMETRI-5][$x = 37$ $\text{mm}$ ($J=75$)]{\label{fig:pdf21}\includegraphics[width=0.25\textwidth]{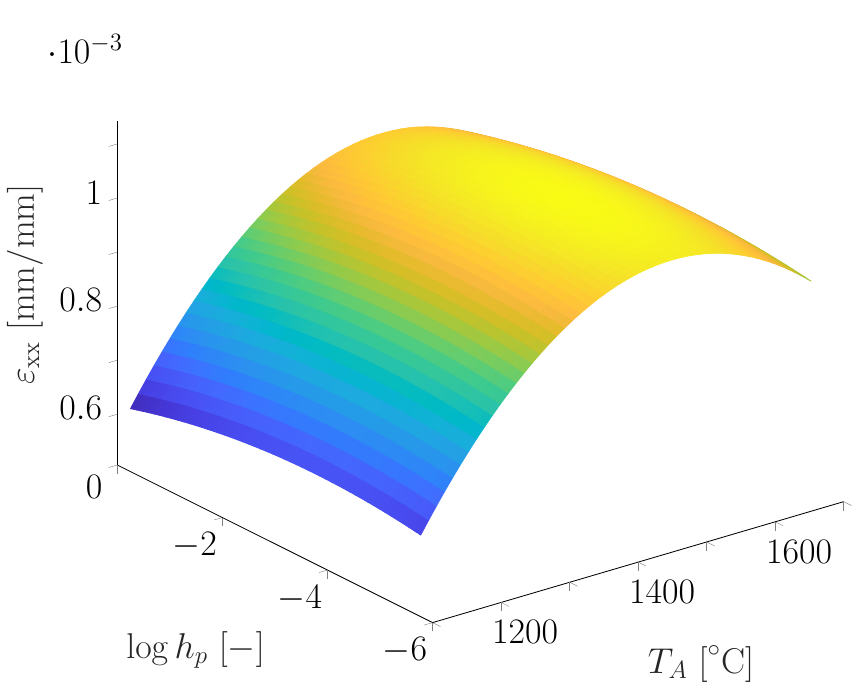}} 
			\subfigure[PARAMETRI-6][$x = 56.5$ $\text{mm}$ ($J=114$)]{\label{fig:pdf22}\includegraphics[width=0.25\textwidth]{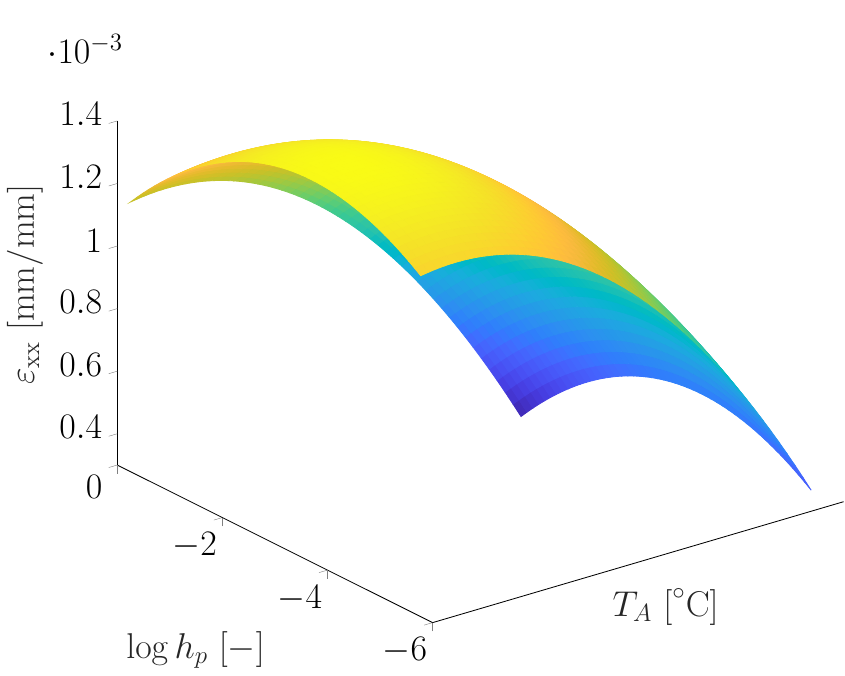}} 
		\end{center}
		\caption{Results of the data-informed forward UQ analysis. MISC multi-fidelity surrogate models
			for some of the nodes of the cantilever beam, obtained from the collocation points shown in \Cref{fig:sparsefor}.}  \label{fig:pdf40}
	\end{figure}

	\Lorenzo{The new set of collocation points is shown in \Cref{fig:sparsefor} and uses the so-called weighted Gaussian Leja points \cite{narayan:leja},
		which are suitable for Gaussian random variables. We initially use only} 
	6 evaluations of the PBF numerical model: 5 evaluations on the low-fidelity model corresponding to $\alpha=1$, and one evaluation on the high-fidelity model corresponding to $\alpha=2$:
	\Lorenzo{we elaborate more on the number of collocation points later on.}
	Examples of the surrogate models constructed for the $J$ beam nodes are shown in \Cref{fig:pdf40}.
	
	
	\subsubsection{Residual strains prediction \Lorenzo{and uncertainty bands}}
	\label{subsubsec:resres}
	
	We are now ready to approximate the residual strains at the $S$ random samples of $T_A$ and $\log h_p$ and compute the corresponding PDFs.
    \rev{For the sake of comparison, we also report the PDFs of the residual strains the can be obtained according to $\rho_{prior}$, i.e., upon building a set of surrogate models for the residual strains based on $\rho_{prior}$ (note that the surrogate models discussed in Section 6.1.2, see, e.g., Figure 6, were for the displacement and not for the residual strains), and evaluating them on a $\rho_{prior}$-based sample of values for $T_A$ and $\log h_p$.
    }
	
\begin{figure}[!ht]
		\begin{center}
			\includegraphics[width=0.6\textwidth]{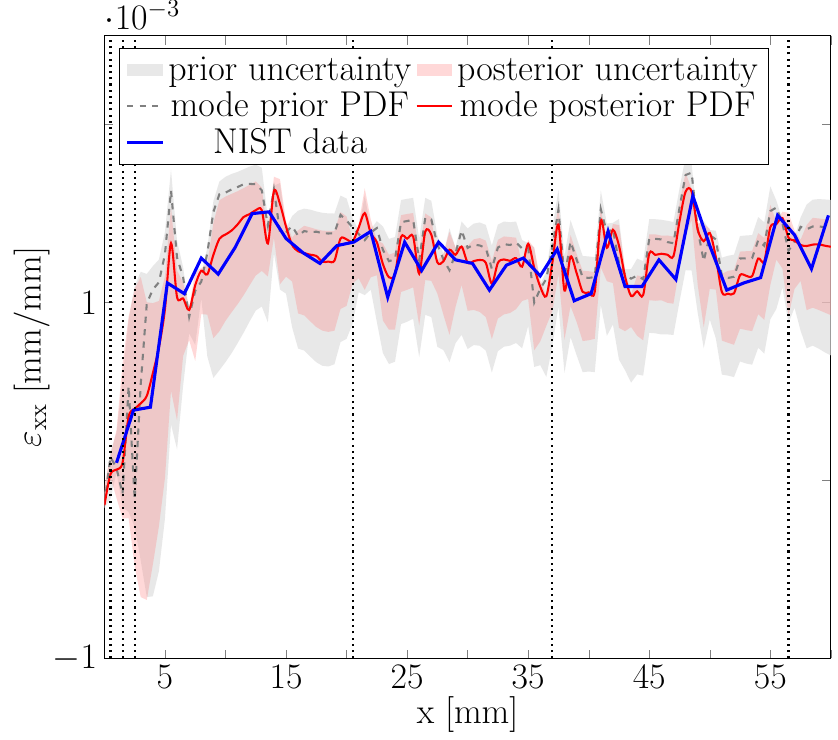}
		\end{center}
		\caption{Results of the data-informed forward UQ analysis. The figure includes:
			(i) the most likely profiles of $\varepsilon_{xx}$ obtained from forward UQ analyses based on
			$\rho_{prior}$ (the dashed grey line represents the the mode of the prior-based PDFs) and
			$\rho_{post}$ (the solid red line represents the mode of the posterior-based PDFs),
			(ii) the associated uncertainty bands, i.e., the 5\% - 95\% quantile bands of the two PDFs
			(grey area for the mode prior-based PDF and pink area for the mode posterior-based PDF) and
			(iii) the residual strains profile provided by NIST (continuous blue line).
			The vertical dotted lines
			represent the 6 positions at which we report the prior and data-informed posterior PDFs in
			\Cref{fig:pdfAA}.}
		\label{fig:strainPred}
	\end{figure}

\begin{figure}[!ht]
		\begin{center}
			\subfigure[PARAMETRI-1][$x = 1.5$ $\text{mm}$]{\label{fig:pdfa}\includegraphics[width=0.3\textwidth]{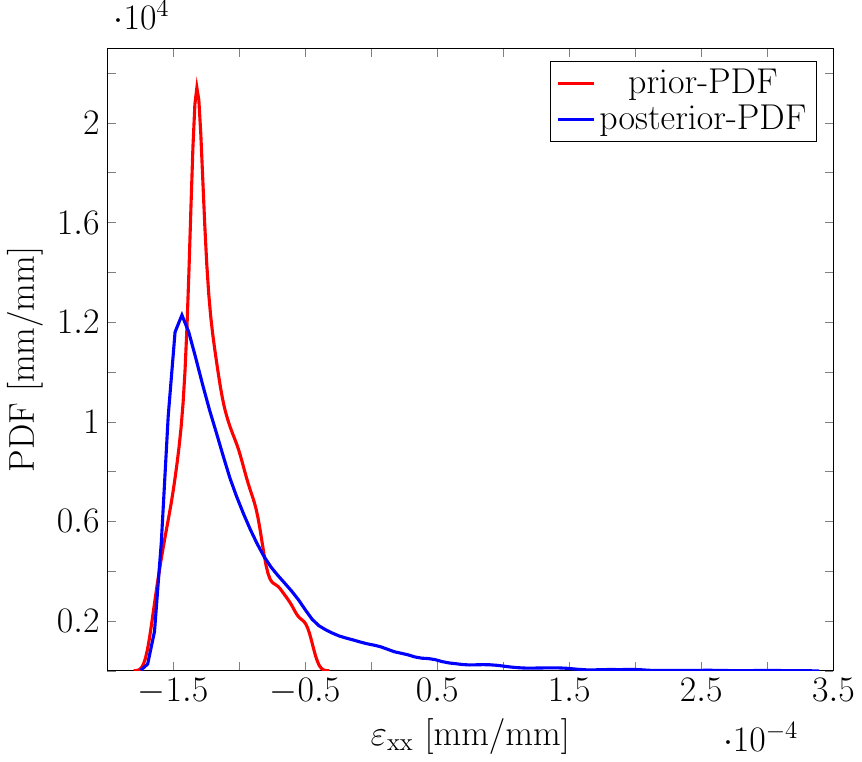}} 
			\subfigure[PARAMETRI-2][$x = 7.5$ $\text{mm}$]{\label{fig:pdfb}\includegraphics[width=0.3\textwidth]{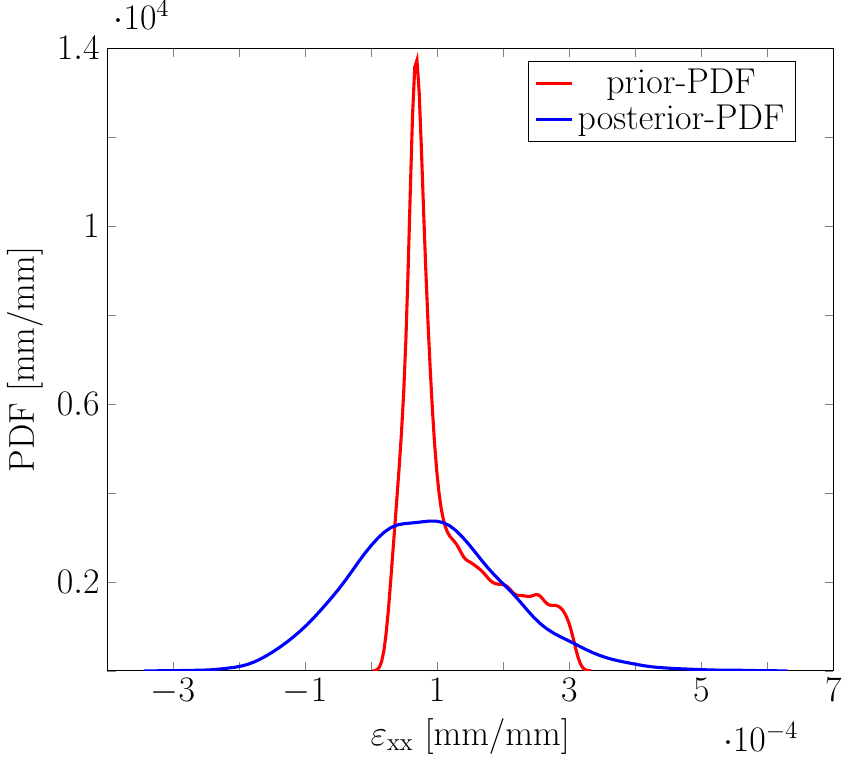}} 
			\subfigure[PARAMETRI-3][$x = 15.5$ $\text{mm}$]{\label{fig:pdfc}\includegraphics[width=0.31\textwidth]{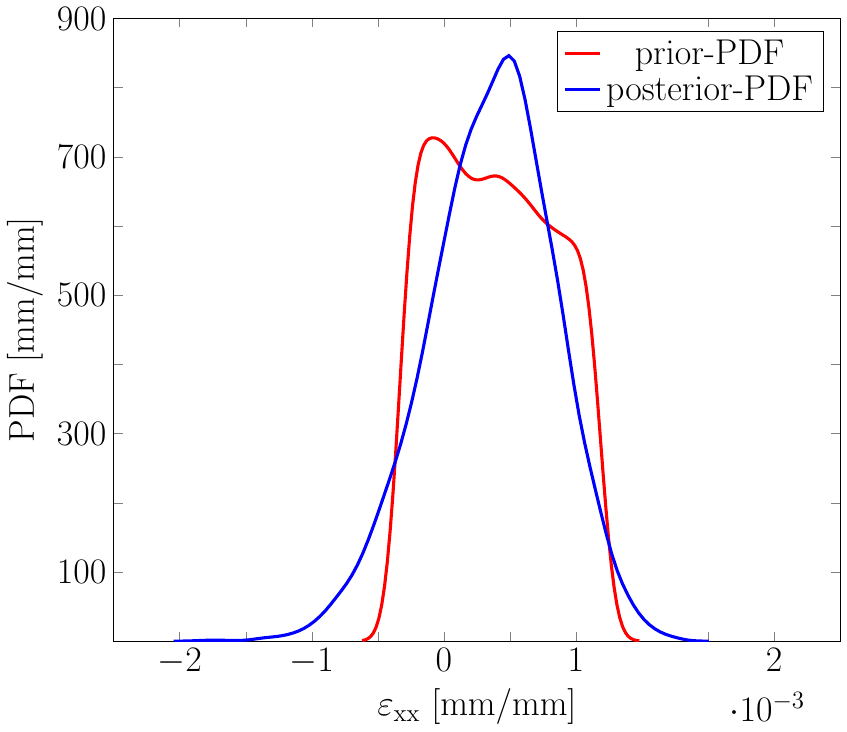}} 
			\subfigure[PARAMETRI-4][$x = 26.5$ $\text{mm}$]{\label{fig:pdfd}\includegraphics[width=0.29\textwidth]{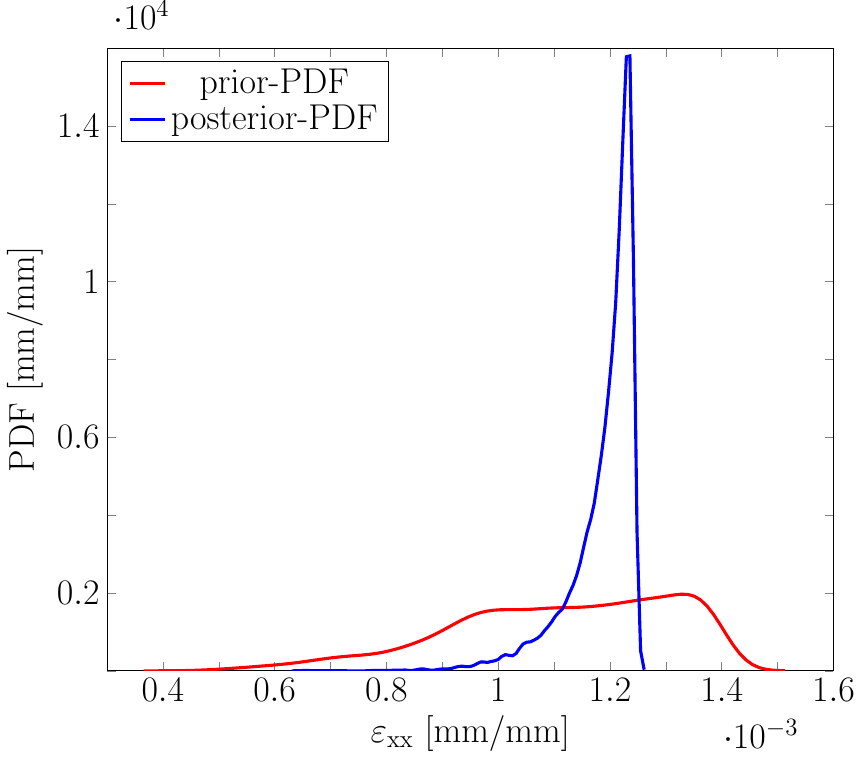}} 
			\subfigure[PARAMETRI-5][$x = 40.5$ $\text{mm}$]{\label{fig:pdfe}\includegraphics[width=0.3\textwidth]{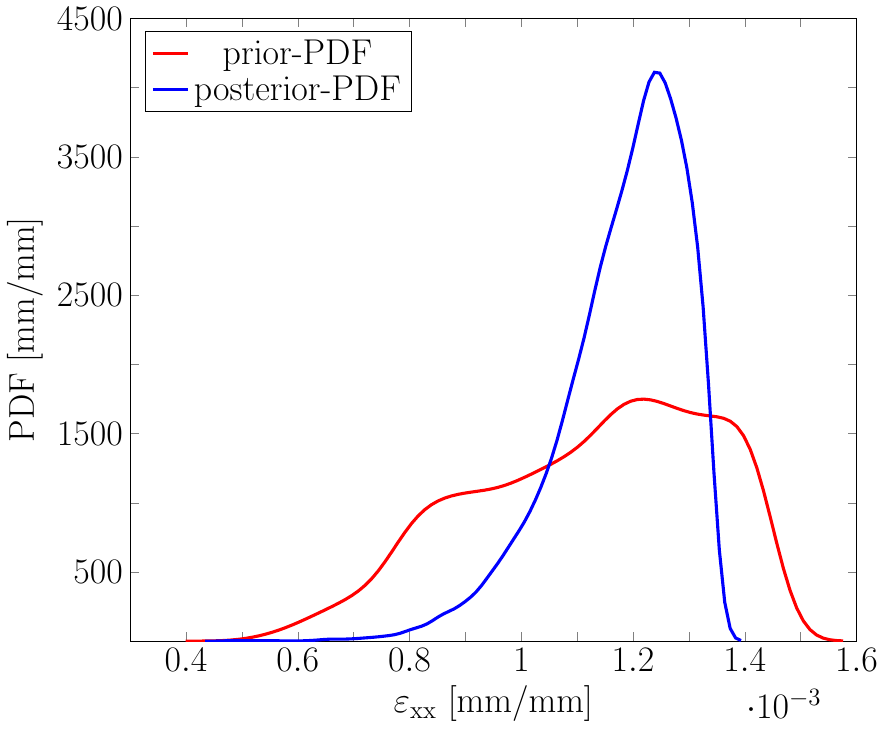}} 
			\subfigure[PARAMETRI-6][$x = 53$ $\text{mm}$]{\label{fig:pdff}\includegraphics[width=0.3\textwidth]{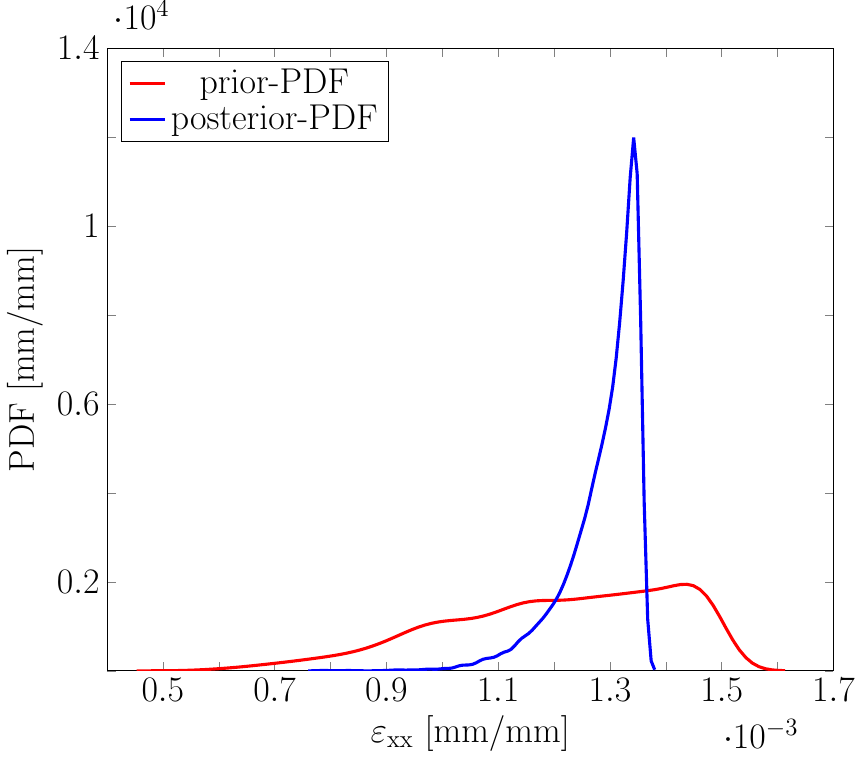}} 
		\end{center}
		\caption{Results of data-informed forward UQ analysis. Prior-based and data-informed PDFs of the residual strains of the cantilever beam
			for the $x$ locations marked in \Cref{fig:dispstrain}.}
		\label{fig:pdfAA}
	\end{figure}

	In \Cref{fig:strainPred} we present the results obtained, which include:
	(i) the most likely profiles of $\varepsilon_{xx}$, \Lorenzo{according to} 
	$\rho_{prior}$ and $\rho_{post}$ respectively, i.e., the modes of the two PDFs of $\varepsilon_{xx}$ at each of the $J=120$ positions
	(the dashed grey line represents the mode of the prior PDFs, while the solid red line represents the mode of  the posterior PDFs),
	and (ii) the associated uncertainty bands, i.e., the 5\% - 95\% quantile bands of the two PDFs
	(grey area for the mode PDF based on the prior PDFs and pink area for the mode posterior-based PDFs).
	\Cref{fig:strainPred} also shows the (iii) the residual strains profile provided by NIST (continuous blue line), while the vertical dotted lines represent the 6 positions at which we report the prior and data-informed PDFs in \Cref{fig:pdfAA}. 
	The residual strains profile provided by NIST overlaps the mode of the posterior-based PDFs,
	indicating that the most likely residual strains profile identified by these PDFs is a faithful representation of the residual strains experimental profile.
	
	\Lorenzo{\Cref{fig:strainPred} also shows that the quantile bands based on the posterior PDFs are less wide than those based on the prior PDFs,
		indicating that overall the Bayesian inversion procedure \rev{was able to significantly
		reduce} the uncertainty in the predictions of the residual strains.}
	To give a more quantitative appraisal of this statement, we compute the quantile band width for the residual strains at each of the $J$ positions using both prior and posterior PDFs. 
	We then average the differences between the two quantile bands, with normalisation relative to the prior quantile band. 
	More precisely, the uncertainty reduction in residual strain predictions is computed as follows:
	\begin{equation*}
		\label{eq:UQ}
		100 \times \frac{1}{J} \sum_{j=1}^{J}  {\frac{{ \Big(Q_{prior,95\%}(\varepsilon_{xx,j})-Q_{prior,5\%}(\varepsilon_{xx,j})\Big)-\Big(Q_{post,95\%}(\varepsilon_{xx,j})-Q_{post,5\%}(\varepsilon_{xx,j})\Big)}}{{Q_{prior,95\%}(\varepsilon_{xx,j})-Q_{prior,5\%}(\varepsilon_{xx,j})}}},
	\end{equation*}
	where $Q_{prior,5\%}$ and $Q_{prior,95\%}$ represent the 5\% and 95\% quantile bands relative to the prior PDFs of  $\varepsilon_{xx}$ for $j=1,...,J$ locations, respectively.  Analogously, $Q_{post,5\%}$ and $Q_{post,95\%}$ are the 5\% and 95\% quantile bands relative to the posterior PDFs of $\varepsilon_{xx}$ for $j=1,...,J$ locations, respectively.
	\Lorenzo{The results indicate a $33\%$ reduction from prior to posterior 
		which is a rather significant improvement.
		Of course, these results are somehow biased by the fact that we willingly took rather large prior intervals for $T_A$ and $\log h_p$
		(cf. \Cref{subsubsec:uncertaindANDQoIs}), i.e., taking narrower prior intervals might have lead to a smaller reduction in the uncertainty.
		On the other hand, the narrower the prior intervals, the higher the risk of missing $\vv_{MAP}$, i.e., the risk that $\vv_{MAP} \not \in \Gamma$
		and thus the Gaussian approximation of the posterior PDF would be centered at the wrong place.
		These results thus overall show that one can safely start from relatively wide prior intervals, as the methodology is quite robust in this respect.}
	
\begin{figure}[!ht]
		\begin{center}
			\subfigure[PARAMETRI-1][]{\label{fig:s1}\includegraphics[width=0.35\textwidth]{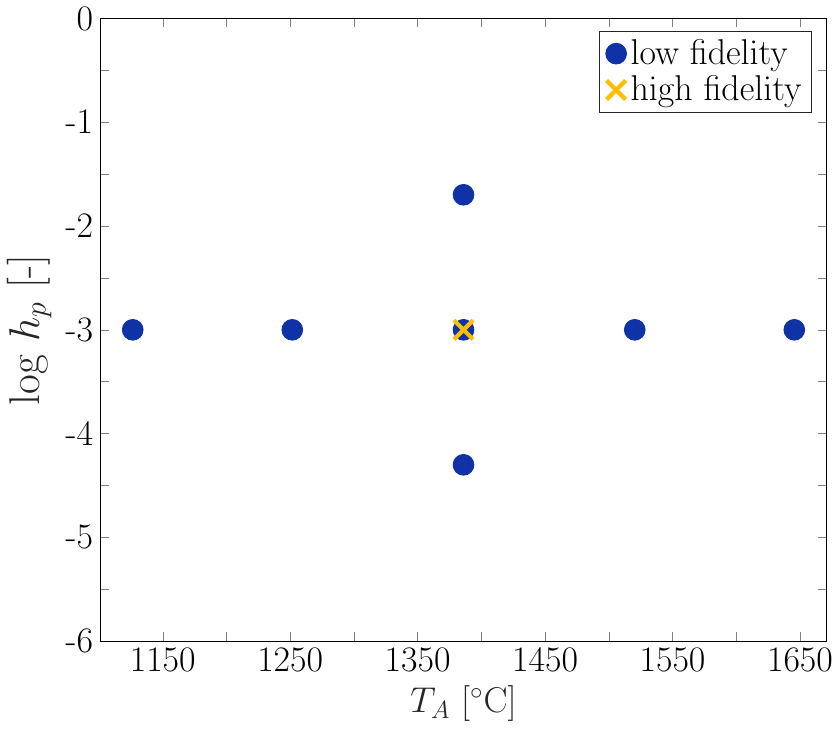}} 
			\subfigure[PARAMETRI-2][]{\label{fig:s1i}\includegraphics[width=0.35\textwidth]{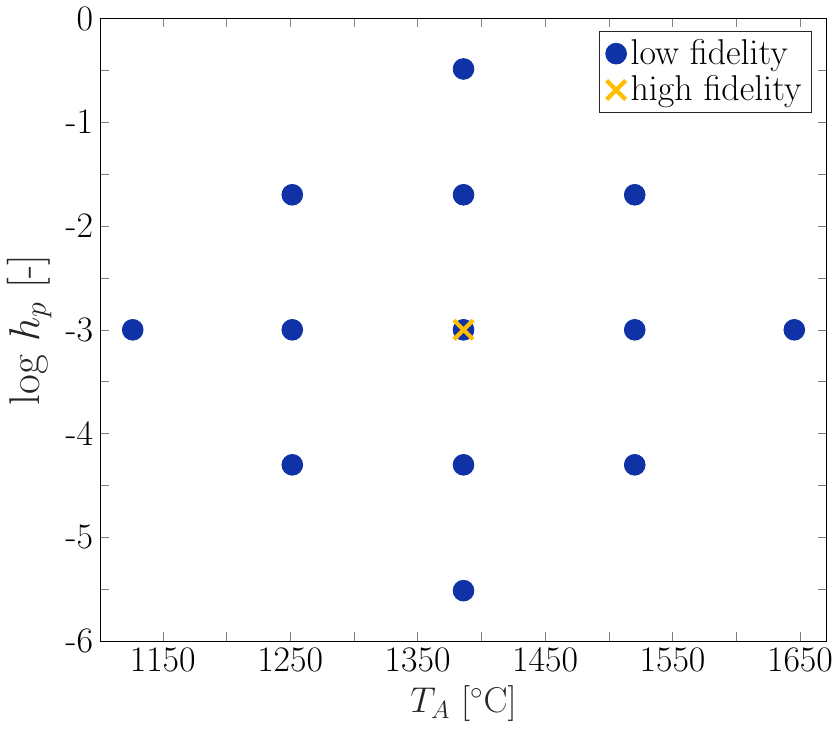}} 
			\subfigure[PARAMETRI-3][]{\label{fig:s2}\includegraphics[width=0.35\textwidth]{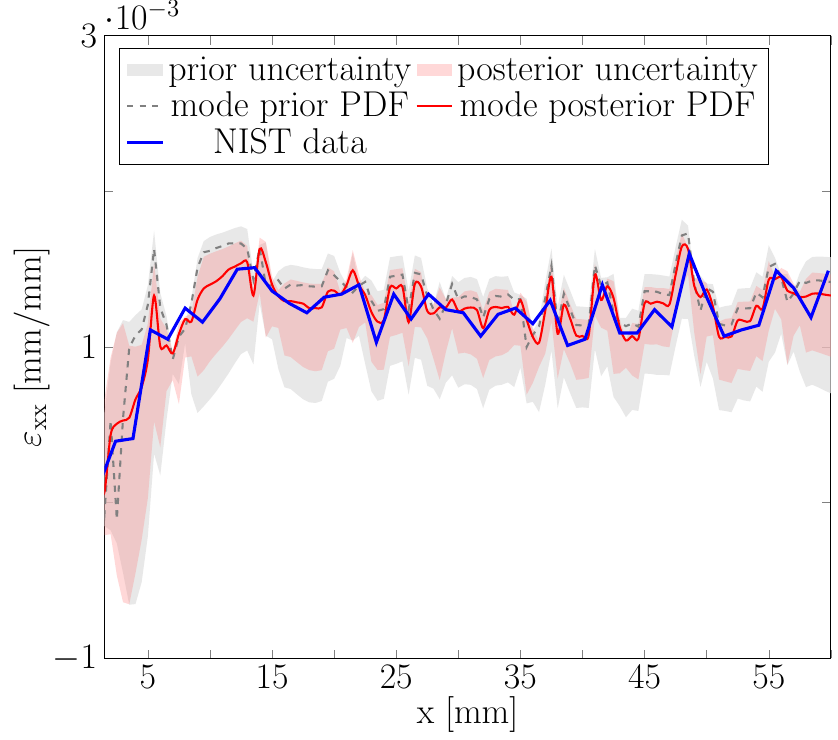}} 
			\subfigure[PARAMETRI-4][]{\label{fig:s3}\includegraphics[width=0.35\textwidth]{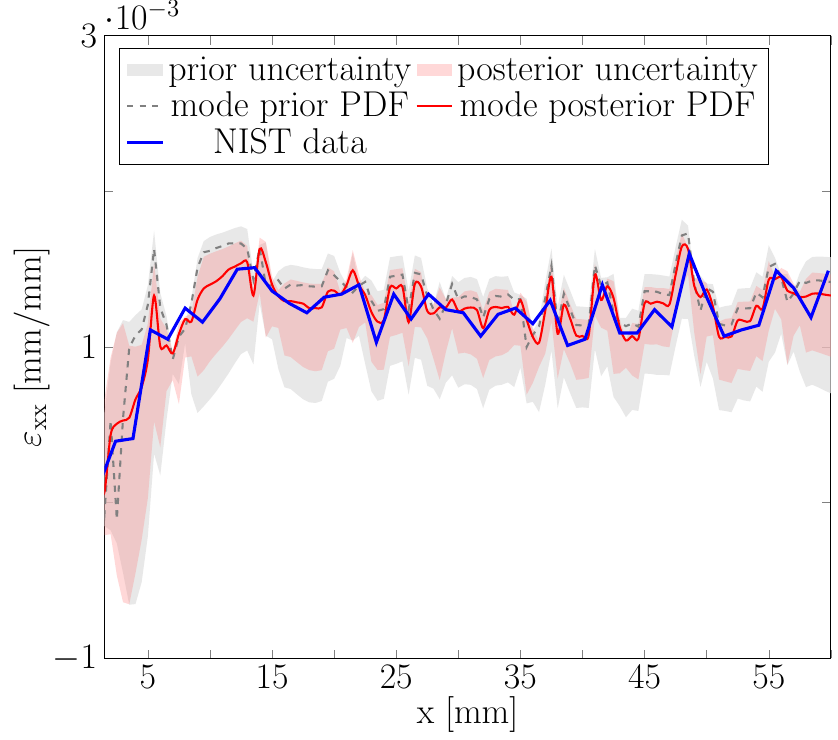}} 
		\end{center}
		\caption{Results of the data-informed forward UQ analysis.
			Top row: collocation points for the MISC surrogate model with 7 (left) and 13 (right) low-fidelity evaluations.
			Bottom row: corresponding modes and uncertainty bands of the residual strains.}
		\label{fig:comp}
	\end{figure}

	\Lorenzo{To further check the validity of the results presented in \Cref{fig:strainPred}, we repeated the forward UQ analysis upon refining
		twice the MISC model and comparing the results thus obtained.
		Specifically, we constructed two new surrogate models by using 7 and 13 evaluations of the low-fidelity model
		(i.e., we added incrementally first 2 collocation points to those in \Cref{fig:sparsefor} and then 6 more,
		given the nestedness of the weighted Leja points -- no high-fidelity evaluations are added), and obtained the results reported in \Cref{fig:comp}.
		Such results show that, despite the increase in the number of evaluations of the PBF-LB/M numerical model,
		there is no significant change, neither in the quantile bands nor in the modes of the PDFs of the residual strains uncertainty.}
	This confirms the validity of choosing the model based on 5 low-fidelity model evaluations and 1 high-fidelity model evaluation,
	as it is not only able to faithfully represent the experimental residual strains, but is also very cost effective from a computational point of view.
	
	\subsubsection{Multi-fidelity versus high-fidelity surrogate models results}
	\label{subsubsec:LowMISCHigh}
	
\begin{figure}[!ht]
		\begin{center}
			%
			\subfigure[PARAMETRI-2][]{\label{fig:s32}\includegraphics[width=0.35\textwidth]{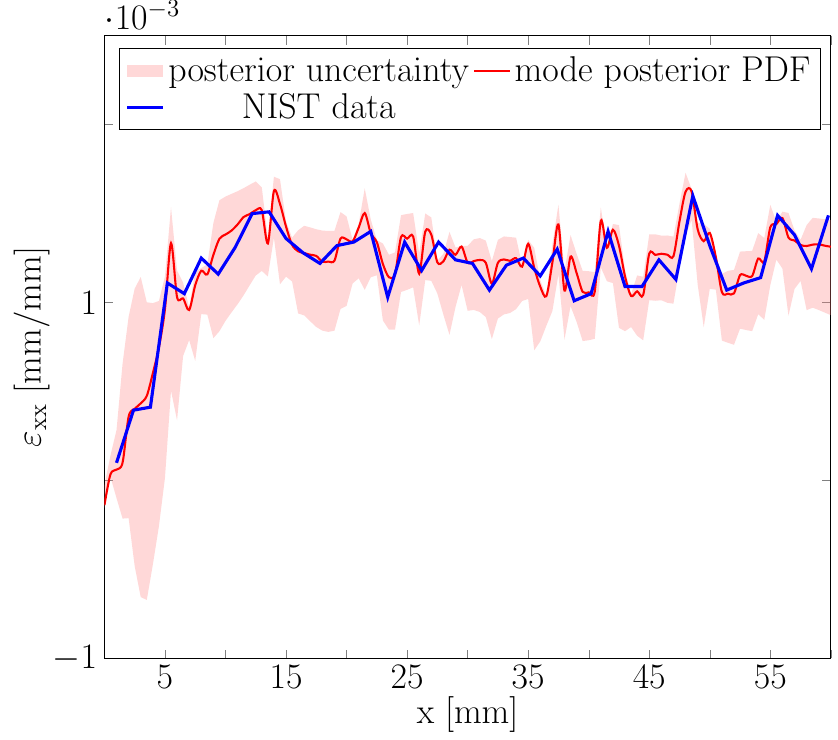}}  
			\subfigure[PARAMETRI-3][]{\label{fig:s33}\includegraphics[width=0.35\textwidth]{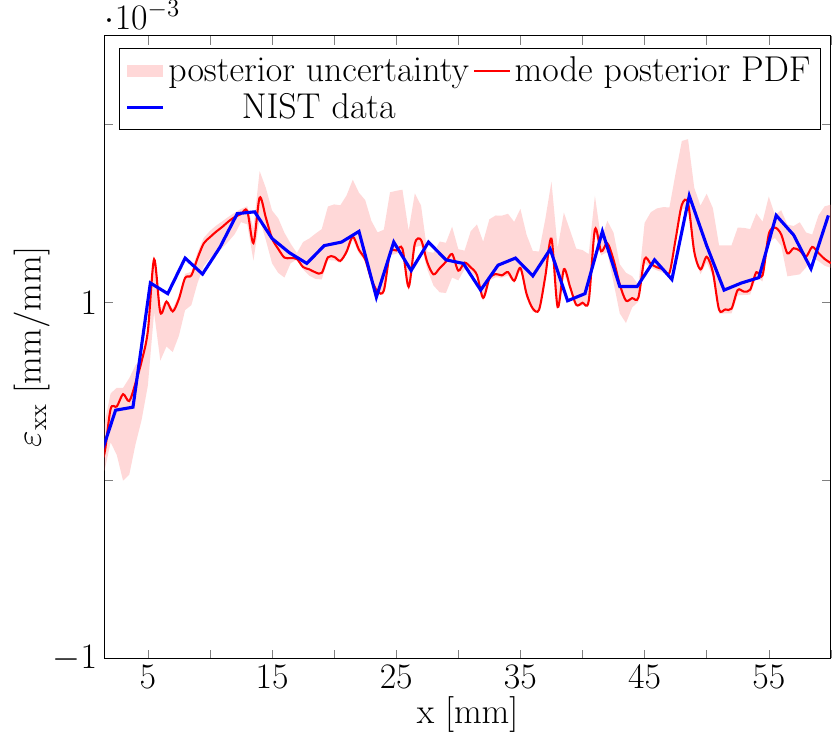}} 
		\end{center}
		\caption{Results of the data-informed forward UQ analysis.
			Comparison in terms of prediction of residual strains of the PBF-LB/M numerical model and associated uncertainty bands using
			\Lorenzo{
				(a) surrogate models based on multi-fidelity PBF-LB/M numerical model evaluations;
				(b) surrogate models based on high-fidelity  PBF-LB/M numerical model evaluations.
			}
		}
		\label{fig:comp2}
	\end{figure}

	To conclude the discussion, we compare the results obtained with the multi-fidelity surrogate model (5 low-fidelity + 1 high-fidelity)
	with respect to those obtained by employing a surrogate model built using high-fidelity simulations only, running 5 high-fidelity simulations at the same collocation points where the multi-fidelity surrogate model requests low-fidelity simulations.
	The results in \Cref{fig:comp2} are similar in terms of the modes for both the multi-fidelity and the high-fidelity surrogate models in the region of $x > 25$ mm, even though the multi-fidelity uncertainty band is below the reference data 
	whereas the high-fidelity band is shifted upward. \rev{Instead}, for $x < 25$ mm the multi-fidelity model shows a larger uncertainty band than the high-fidelity model. A possible explanation is that since this region is more critical in terms of residual strains, low-fidelity simulations are not sufficiently accurate. However, it is worth to remark that the computational time required to run 5 high-fidelity simulations is 36 times larger than the time of 5 low-fidelity analyses (see \Cref{tab:FE}). \rev{Therefore, it is questionable whether such extra effort is actually worth it.}
	\section{Conclusions}
	\label{sec:Conclusion}
	
	In the present paper, we address the complex challenge of uncertainty quantification in the context of numerical simulations of the PBF-LB/M
	process with remarkable success, focusing on a part-scale thermomechanical model of an Inconel 625 beam.
	Our innovative methodology combines part-scale thermomechanical numerical simulations with MISC multi-fidelity surrogate models and UQ techniques.
	
	First, we performed a Bayesian inverse UQ analysis \Lorenzo{based on experimental data} to evaluate uncertainties associated with simulation parameters, 
	\Lorenzo{namely the} powder convection coefficient and activation temperature.
	Afterwards, we applied a data-informed forward UQ analysis to predict residual strains in the cantilever beam, by 
	\Lorenzo{computing approximations of} the PDFs.
	Our methodology efficiently exploited MISC surrogate models, greatly reducing the computational burden of the numerous simulations required by the UQ analysis. The entire procedure, from Bayesian inverse UQ to data-informed forward UQ analysis, required only 28 numerical simulations of the PBF-LB/M model, which highlights the effectiveness and efficiency of our surrogate modelling approach.
	
	\Lorenzo{The results show} that the predictions based on the data-informed PDF are significantly more accurate than those based on the \Lorenzo{prior PDF
		(even though we were admittedly conservative in choosing the prior intervals for the two uncertain parameters).
		In particular, we observe a reduction on the average standard deviations of the residual strains of $33\%$,} 
	with our residual strain predictions in strict agreement with the experimental data provided by NIST.

	\section*{Author contributions}
	
	\textbf{Mihaela Chiappetta:} Software, Visualisation, Writing – original draft. \textbf{Chiara Piazzola:} Formal analysis, Software, Writing – review \& editing. \textbf{Lorenzo Tamellini:} Formal analysis, Methodology, Writing – review \& editing. \textbf{Alessandro Reali:} Supervision, Funding acquisition. \textbf{Ferdinando Auricchio:} Supervision, Funding acquisition. \textbf{Massimo Carraturo:} Supervision, Conceptualisation, Writing – review \& editing.
	
	\section*{Acknowledgments}
	This work was partially supported by the Italian Minister of University and Research through the MIUR-PRIN projects
	\lq\lq A BRIDGE TO THE FUTURE: Computational methods, innovative applications, experimental validations of new materials and technologies\rq\rq~(No. 2017L7X3CS) and
	\lq\lq XFAST-SIMS\rq\rq~(No. 20173C478N).
	\mihi{This work was partially supported by the Italian Ministry of the University and Research (MUR) through the FSE – REACT EU: PON Ricerca e Innovazione 2014-2020.}
	Lorenzo Tamellini is member of the Gruppo Nazionale Calcolo Scientifico-Istituto Nazionale di Alta Matematica (GNCS-INdAM) and has been supported by the PRIN 2022 project ``Numerical approximation of uncertainty quantification problems for PDEs by multi-fidelity methods (UQ-FLY)'' (No. 202222PACR), funded by the European Union - NextGenerationEU.
    Chiara Piazzola is member of the Gruppo Nazionale Calcolo Scientifico-Istituto Nazionale di Alta Matematica (GNCS-INdAM) and acknowledges the support of the Alexander von Humboldt Foundation. 
	Mihaela Chiappetta, Massimo Carraturo and Ferdinando Auricchio acknowledge partial financial support from the Ministry of Enterprise and Made in Italy and Lombardy Region through the project \lq\lq PRotesi innOvaTivE per applicazioni vaScolari ed ortopedIChe e mediante Additive Manufacturing\rq\rq - CUP : B19J22002460005, on the finance  Asse I, Azione 1.1.3 PON Businesses and Competitiveness 2014 - 2020.
	{Mihaela Chiappetta, Lorenzo Tamellini, Alessandro Reali, and Ferdinando Auricchio have been partially supported by ICSC -- Centro Nazionale di Ricerca in High Performance Computing, Big Data, and Quantum Computing funded by European Union -- NextGenerationEU}.
	
	%
	
	\section*{Conflict of interest}
	
	The authors declare that they have no known competing financial interests or personal relationships that could have appeared to influence the work reported in this paper.

	\begin{figure*}[!ht]
	\begin{center}
	   \subfigure{\includegraphics[width=0.25\textwidth]{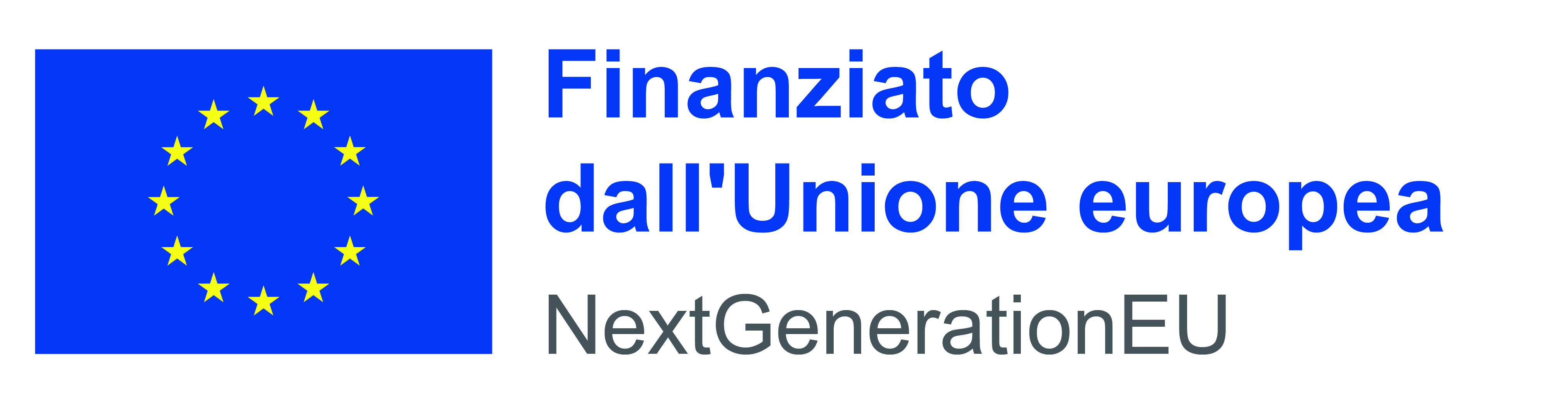}} \hspace{1cm}
		\subfigure{\includegraphics[width=0.2\textwidth]{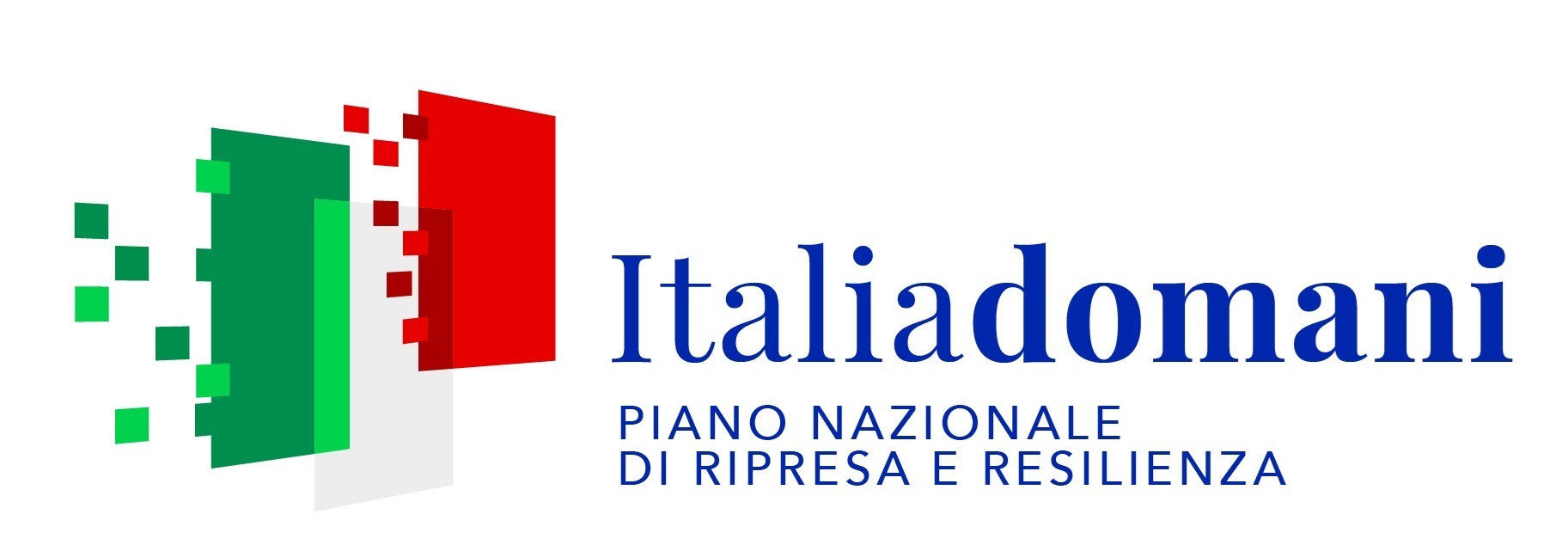}}  \hspace{1.3cm}
		\subfigure{\includegraphics[width=0.1\textwidth]{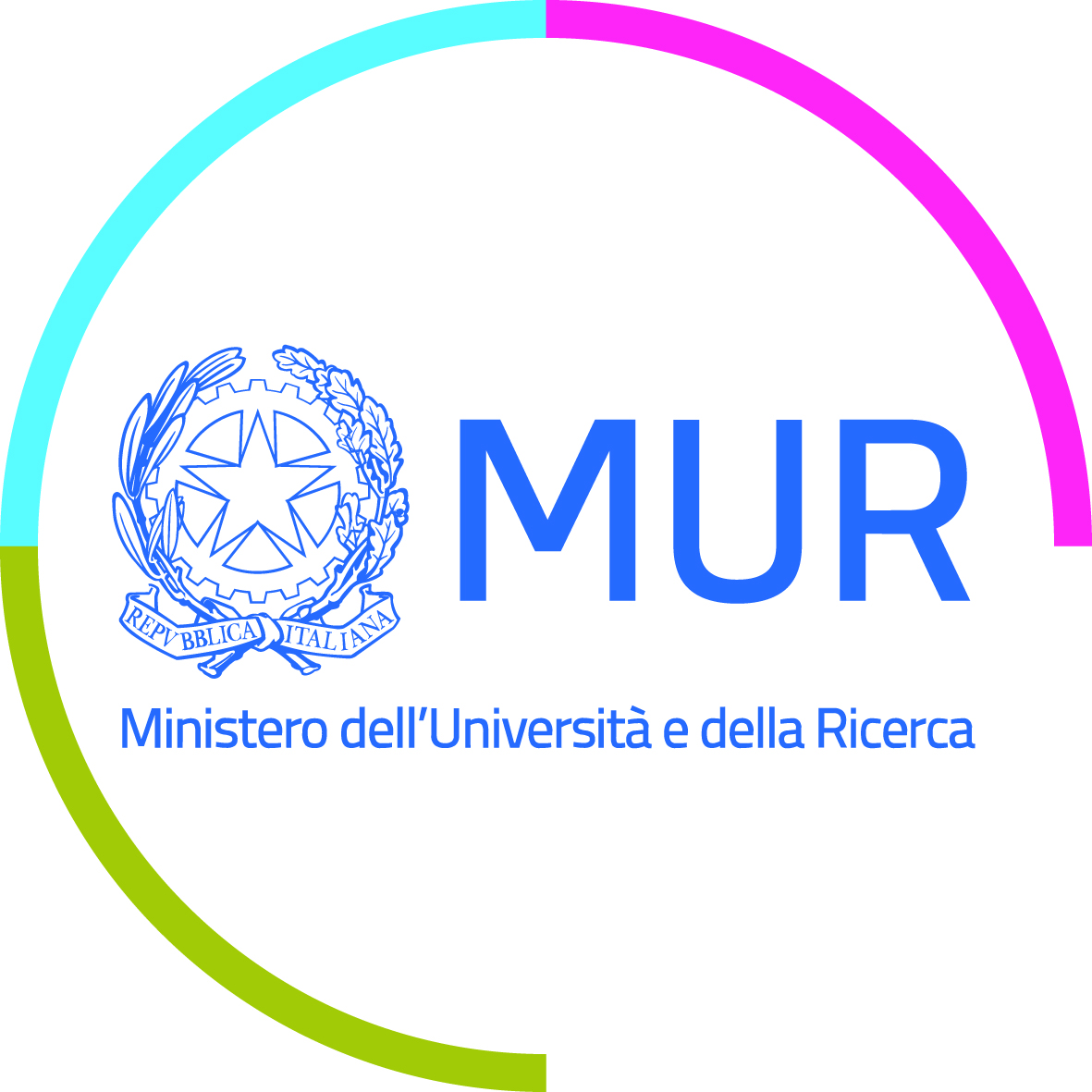}} 
	\end{center}
\end{figure*}	
	
\bibliographystyle{unsrtnat}
\bibliography{ReferenceMISC.bbl}

\end{document}